\documentclass{article}
\usepackage[a4paper, total={7in, 10in}]{geometry}
\usepackage[table]{xcolor}
\definecolor{lightgray}{gray}{0.9}
\usepackage[utf8]{inputenc}
\usepackage[hyphens]{url}
\usepackage{booktabs}
\usepackage{subcaption}
\usepackage[ruled,vlined]{algorithm2e}
\usepackage[noend]{algpseudocode}
\definecolor{lst-gray}{rgb}{0.98,0.98,0.98}
\definecolor{lst-blue}{RGB}{40,0.0,255}
\definecolor{lst-green}{RGB}{65,128,95}
\definecolor{lst-red}{RGB}{200,0,85}

\usepackage{listings}
\lstdefinelanguage{PowerShell}{
	morekeywords={
		Add-Content,Add-PSSnapin,Clear-Content,Clear-History,Clear-Host,Clear-Item,Clear-ItemProperty,Clear-Variable,Compare-Object,Connect-PSSession,ConvertFrom-String,Convert-Path,Copy-Item,Copy-ItemProperty,Disable-PSBreakpoint,Disconnect-PSSession,Enable-PSBreakpoint,Enter-PSSession,Exit-PSSession,Export-Alias,Export-Csv,Export-PSSession,ForEach-Object,Format-Custom,Format-Hex,Format-List,Format-Table,Format-Wide,Get-Alias,Get-ChildItem,Get-Clipboard,Get-Command,Get-ComputerInfo,Get-Content,Get-History,Get-Item,Get-ItemProperty,Get-ItemPropertyValue,Get-Job,Get-Location,Get-Member,Get-Module,Get-Process,Get-PSBreakpoint,Get-PSCallStack,Get-PSDrive,Get-PSSession,Get-PSSnapin,Get-Service,Get-TimeZone,Get-Unique,Get-Variable,Get-WmiObject,Group-Object,help,Import-Alias,Import-Csv,Import-Module,Import-PSSession,Invoke-Command,Invoke-Expression,Invoke-History,Invoke-Item,Invoke-RestMethod,Invoke-WebRequest,Invoke-WmiMethod,Measure-Object,mkdir,Move-Item,Move-ItemProperty,New-Alias,New-Item,New-Module,New-PSDrive,New-PSSession,New-PSSessionConfigurationFile,New-Variable,Out-GridView,Out-Host,Out-Printer,Pop-Location,powershell_ise.exe,Push-Location,Receive-Job,Receive-PSSession,Remove-Item,Remove-ItemProperty,Remove-Job,Remove-Module,Remove-PSBreakpoint,Remove-PSDrive,Remove-PSSession,Remove-PSSnapin,Remove-Variable,Remove-WmiObject,Rename-Item,Rename-ItemProperty,Resolve-Path,Resume-Job,Select-Object,Select-String,Set-Alias,Set-Clipboard,Set-Content,Set-Item,Set-ItemProperty,Set-Location,Set-PSBreakpoint,Set-TimeZone,Set-Variable,Set-WmiInstance,Show-Command,Sort-Object,Start-Job,Start-Process,Start-Service,Start-Sleep,Stop-Job,Stop-Process,Stop-Service,Suspend-Job,Tee-Object,Trace-Command,Wait-Job,Where-Object,Write-Output
	},
	morekeywords={
		Add-AppxPackage,Add-AppxProvisionedPackage,Add-AppxVolume,Add-BitsFile,Add-CertificateEnrollmentPolicyServer,Add-Computer,Add-Content,Add-History,Add-JobTrigger,Add-KdsRootKey,Add-LocalGroupMember,Add-Member,Add-PSSnapin,Add-Type,Add-WindowsCapability,Add-WindowsDriver,Add-WindowsImage,Add-WindowsPackage,Checkpoint-Computer,Clear-Content,Clear-EventLog,Clear-History,Clear-Item,Clear-ItemProperty,Clear-KdsCache,Clear-RecycleBin,Clear-Tpm,Clear-Variable,Clear-WindowsCorruptMountPoint,Compare-Object,Complete-BitsTransfer,Complete-DtiagnosticTransaction,Complete-Transaction,Confirm-SecureBootUEFI,Connect-PSSession,Connect-WSMan,ConvertFrom-Csv,ConvertFrom-Json,ConvertFrom-SecureString,ConvertFrom-String,ConvertFrom-StringData,Convert-Path,Convert-String,ConvertTo-Csv,ConvertTo-Html,ConvertTo-Json,ConvertTo-ProcessMitigationPolicy,ConvertTo-SecureString,ConvertTo-TpmOwnerAuth,ConvertTo-Xml,Copy-Item,Copy-ItemProperty,Debug-Job,Debug-Process,Debug-Runspace,Disable-AppBackgroundTaskDiagnosticLog,Disable-ComputerRestore,Disable-JobTrigger,Disable-LocalUser,Disable-PSBreakpoint,Disable-PSRemoting,Disable-PSSessionConfiguration,Disable-RunspaceDebug,Disable-ScheduledJob,Disable-TlsCipherSuite,Disable-TlsEccCurve,Disable-TlsSessionTicketKey,Disable-TpmAutoProvisioning,Disable-WindowsErrorReporting,Disable-WindowsOptionalFeature,Disable-WSManCredSSP,Disconnect-PSSession,Disconnect-WSMan,Dismount-AppxVolume,Dismount-WindowsImage,Enable-AppBackgroundTaskDiagnosticLog,Enable-ComputerRestore,Enable-JobTrigger,Enable-LocalUser,Enable-PSBreakpoint,Enable-PSRemoting,Enable-PSSessionConfiguration,Enable-RunspaceDebug,Enable-ScheduledJob,Enable-TlsCipherSuite,Enable-TlsEccCurve,Enable-TlsSessionTicketKey,Enable-TpmAutoProvisioning,Enable-WindowsErrorReporting,Enable-WindowsOptionalFeature,Enable-WSManCredSSP,Enter-PSHostProcess,Enter-PSSession,Exit-PSHostProcess,Exit-PSSession,Expand-WindowsCustomDataImage,Expand-WindowsImage,Export-Alias,Export-BinaryMiLog,Export-Certificate,Export-Clixml,Export-Console,Export-Counter,Export-Csv,Export-FormatData,Export-ModuleMember,Export-PfxCertificate,Export-ProvisioningPackage,Export-PSSession,Export-StartLayout,Export-StartLayoutEdgeAssets,Export-TlsSessionTicketKey,Export-Trace,Export-WindowsCapabilitySource,Export-WindowsDriver,Export-WindowsImage,Find-Package,Find-PackageProvider,ForEach-Object,Format-Custom,Format-List,Format-SecureBootUEFI,Format-Table,Format-Wide,Get-Acl,Get-Alias,Get-AppxDefaultVolume,Get-AppxPackage,Get-AppxPackageManifest,Get-AppxProvisionedPackage,Get-AppxVolume,Get-AuthenticodeSignature,Get-BitsTransfer,Get-Certificate,Get-CertificateAutoEnrollmentPolicy,Get-CertificateEnrollmentPolicyServer,Get-CertificateNotificationTask,Get-ChildItem,Get-CimAssociatedInstance,Get-CimClass,Get-CimInstance,Get-CimSession,Get-Clipboard,Get-CmsMessage,Get-Command,Get-ComputerInfo,Get-ComputerRestorePoint,Get-Content,Get-ControlPanelItem,Get-Counter,Get-Credential,Get-Culture,Get-DAPolicyChange,Get-Date,Get-DeliveryOptimizationLog,Get-DeliveryOptimizationPerfSnap,Get-DeliveryOptimizationPerfSnapThisMonth,Get-DeliveryOptimizationStatus,Get-DODownloadMode,Get-DOPercentageMaxBackgroundBandwidth,Get-DOPercentageMaxForegroundBandwidth,Get-Event,Get-EventLog,Get-EventSubscriber,Get-ExecutionPolicy,Get-FormatData,Get-Help,Get-History,Get-Host,Get-HotFix,Get-Item,Get-ItemProperty,Get-ItemPropertyValue,Get-Job,Get-JobTrigger,Get-KdsConfiguration,Get-KdsRootKey,Get-LocalGroup,Get-LocalGroupMember,Get-LocalUser,Get-Location,Get-Member,Get-Module,Get-Package,Get-PackageProvider,Get-PackageSource,Get-PfxCertificate,Get-PfxData,Get-PmemDisk,Get-PmemPhysicalDevice,Get-PmemUnusedRegion,Get-Process,Get-ProcessMitigation,Get-ProvisioningPackage,Get-PSBreakpoint,Get-PSCallStack,Get-PSDrive,Get-PSHostProcessInfo,Get-PSProvider,Get-PSReadlineKeyHandler,Get-PSReadlineOption,Get-PSSession,Get-PSSessionCapability,Get-PSSessionConfiguration,Get-PSSnapin,Get-Random,Get-Runspace,Get-RunspaceDebug,Get-ScheduledJob,Get-ScheduledJobOption,Get-SecureBootPolicy,Get-SecureBootUEFI,Get-Service,Get-TimeZone,Get-TlsCipherSuite,Get-TlsEccCurve,Get-Tpm,Get-TpmEndorsementKeyInfo,Get-TpmSupportedFeature,Get-TraceSource,Get-Transaction,Get-TroubleshootingPack,Get-TrustedProvisioningCertificate,Get-TypeData,Get-UICulture,Get-Unique,Get-Variable,Get-WIMBootEntry,Get-WinAcceptLanguageFromLanguageListOptOut,Get-WinCultureFromLanguageListOptOut,Get-WinDefaultInputMethodOverride,Get-WindowsCapability,Get-WindowsDeveloperLicense,Get-WindowsDriver,Get-WindowsEdition,Get-WindowsErrorReporting,Get-WindowsImage,Get-WindowsImageContent,Get-WindowsOptionalFeature,Get-WindowsPackage,Get-WindowsSearchSetting,Get-WinEvent,Get-WinHomeLocation,Get-WinLanguageBarOption,Get-WinSystemLocale,Get-WinUILanguageOverride,Get-WinUserLanguageList,Get-WmiObject,Get-WSManCredSSP,Get-WSManInstance,Group-Object,Import-Alias,Import-BinaryMiLog,Import-Certificate,Import-Clixml,Import-Counter,Import-Csv,Import-LocalizedData,Import-Module,Import-PackageProvider,Import-PfxCertificate,Import-PSSession,Import-StartLayout,Import-TpmOwnerAuth,Initialize-PmemPhysicalDevice,Initialize-Tpm,Install-Package,Install-PackageProvider,Install-ProvisioningPackage,Install-TrustedProvisioningCertificate,Invoke-CimMethod,Invoke-Command,Invoke-CommandInDesktopPackage,Invoke-DscResource,Invoke-Expression,Invoke-History,Invoke-Item,Invoke-RestMethod,Invoke-TroubleshootingPack,Invoke-WebRequest,Invoke-WmiMethod,Invoke-WSManAction,Join-DtiagnosticResourceManager,Join-Path,Limit-EventLog,Measure-Command,Measure-Object,Mount-AppxVolume,Mount-WindowsImage,Move-AppxPackage,Move-Item,Move-ItemProperty,New-Alias,New-CertificateNotificationTask,New-CimInstance,New-CimSession,New-CimSessionOption,New-DtiagnosticTransaction,New-Event,New-EventLog,New-FileCatalog,New-Item,New-ItemProperty,New-JobTrigger,New-LocalGroup,New-LocalUser,New-Module,New-ModuleManifest,New-NetIPsecAuthProposal,New-NetIPsecMainModeCryptoProposal,New-NetIPsecQuickModeCryptoProposal,New-Object,New-PmemDisk,New-ProvisioningRepro,New-PSDrive,New-PSRoleCapabilityFile,New-PSSession,New-PSSessionConfigurationFile,New-PSSessionOption,New-PSTransportOption,New-PSWorkflowExecutionOption,New-ScheduledJobOption,New-SelfSignedCertificate,New-Service,New-TimeSpan,New-TlsSessionTicketKey,New-Variable,New-WebServiceProxy,New-WindowsCustomImage,New-WindowsImage,New-WinEvent,New-WinUserLanguageList,New-WSManInstance,New-WSManSessionOption,Optimize-AppxProvisionedPackages,Optimize-WindowsImage,Out-Default,Out-File,Out-GridView,Out-Host,Out-Null,Out-Printer,Out-String,Pop-Location,Protect-CmsMessage,Publish-DscConfiguration,Push-Location,Read-Host,Receive-DtiagnosticTransaction,Receive-Job,Receive-PSSession,Register-ArgumentCompleter,Register-CimIndicationEvent,Register-EngineEvent,Register-ObjectEvent,Register-PackageSource,Register-PSSessionConfiguration,Register-ScheduledJob,Register-WmiEvent,Remove-AppxPackage,Remove-AppxProvisionedPackage,Remove-AppxVolume,Remove-BitsTransfer,Remove-CertificateEnrollmentPolicyServer,Remove-CertificateNotificationTask,Remove-CimInstance,Remove-CimSession,Remove-Computer,Remove-Event,Remove-EventLog,Remove-Item,Remove-ItemProperty,Remove-Job,Remove-JobTrigger,Remove-LocalGroup,Remove-LocalGroupMember,Remove-LocalUser,Remove-Module,Remove-PmemDisk,Remove-PSBreakpoint,Remove-PSDrive,Remove-PSReadlineKeyHandler,Remove-PSSession,Remove-PSSnapin,Remove-TypeData,Remove-Variable,Remove-WindowsCapability,Remove-WindowsDriver,Remove-WindowsImage,Remove-WindowsPackage,Remove-WmiObject,Remove-WSManInstance,Rename-Computer,Rename-Item,Rename-ItemProperty,Rename-LocalGroup,Rename-LocalUser,Repair-WindowsImage,Reset-ComputerMachinePassword,Resolve-DnsName,Resolve-Path,Restart-Computer,Restart-Service,Restore-Computer,Resume-BitsTransfer,Resume-Job,Resume-ProvisioningSession,Resume-Service,Save-Help,Save-Package,Save-WindowsImage,Select-Object,Select-String,Select-Xml,Send-DtiagnosticTransaction,Send-MailMessage,Set-Acl,Set-Alias,Set-AppBackgroundTaskResourcePolicy,Set-AppxDefaultVolume,Set-AppXProvisionedDataFile,Set-AuthenticodeSignature,Set-BitsTransfer,Set-CertificateAutoEnrollmentPolicy,Set-CimInstance,Set-Clipboard,Set-Content,Set-Culture,Set-Date,Set-DODownloadMode,Set-DOPercentageMaxBackgroundBandwidth,Set-DOPercentageMaxForegroundBandwidth,Set-DscLocalConfigurationManager,Set-ExecutionPolicy,Set-Item,Set-ItemProperty,Set-JobTrigger,Set-KdsConfiguration,Set-LocalGroup,Set-LocalUser,Set-Location,Set-PackageSource,Set-ProcessMitigation,Set-PSBreakpoint,Set-PSDebug,Set-PSReadlineKeyHandler,Set-PSReadlineOption,Set-PSSessionConfiguration,Set-ScheduledJob,Set-ScheduledJobOption,Set-SecureBootUEFI,Set-Service,Set-StrictMode,Set-TimeZone,Set-TpmOwnerAuth,Set-TraceSource,Set-Variable,Set-WinAcceptLanguageFromLanguageListOptOut,Set-WinCultureFromLanguageListOptOut,Set-WinDefaultInputMethodOverride,Set-WindowsEdition,Set-WindowsProductKey,Set-WindowsSearchSetting,Set-WinHomeLocation,Set-WinLanguageBarOption,Set-WinSystemLocale,Set-WinUILanguageOverride,Set-WinUserLanguageList,Set-WmiInstance,Set-WSManInstance,Set-WSManQuickConfig,Show-Command,Show-ControlPanelItem,Show-EventLog,Show-WindowsDeveloperLicenseRegistration,Sort-Object,Split-Path,Split-WindowsImage,Start-BitsTransfer,Start-DscConfiguration,Start-DtiagnosticResourceManager,Start-Job,Start-Process,Start-Service,Start-Sleep,Start-Transaction,Start-Transcript,Stop-Computer,Stop-DtiagnosticResourceManager,Stop-Job,Stop-Process,Stop-Service,Stop-Transcript,Suspend-BitsTransfer,Suspend-Job,Suspend-Service,Switch-Certificate,Tee-Object,Test-Certificate,Test-ComputerSecureChannel,Test-Connection,Test-DscConfiguration,Test-FileCatalog,Test-KdsRootKey,Test-ModuleManifest,Test-Path,Test-PSSessionConfigurationFile,Test-WSMan,Trace-Command,Unblock-File,Unblock-Tpm,Undo-DtiagnosticTransaction,Undo-Transaction,Uninstall-Package,Uninstall-ProvisioningPackage,Uninstall-TrustedProvisioningCertificate,Unprotect-CmsMessage,Unregister-Event,Unregister-PackageSource,Unregister-PSSessionConfiguration,Unregister-ScheduledJob,Unregister-WindowsDeveloperLicense,Update-FormatData,Update-Help,Update-List,Update-TypeData,Update-WIMBootEntry,Use-Transaction,Use-WindowsUnattend,Wait-Debugger,Wait-Event,Wait-Job,Wait-Process,Where-Object,Write-Debug,Write-Error,Write-EventLog,Write-Host,Write-Information,Write-Output,Write-Progress,Write-Verbose,Write-Warning
	},
	morekeywords={
		Add-BitLockerKeyProtector,Add-DnsClientNrptRule,Add-DtcClusterTMMapping,Add-EtwTraceProvider,Add-InitiatorIdToMaskingSet,Add-MpPreference,Add-NetEventNetworkAdapter,Add-NetEventPacketCaptureProvider,Add-NetEventProvider,Add-NetEventVFPProvider,Add-NetEventVmNetworkAdapter,Add-NetEventVmSwitch,Add-NetEventVmSwitchProvider,Add-NetEventWFPCaptureProvider,Add-NetIPHttpsCertBinding,Add-NetLbfoTeamMember,Add-NetLbfoTeamNic,Add-NetNatExternalAddress,Add-NetNatStaticMapping,Add-NetSwitchTeamMember,Add-Odbsn,Add-PartitionAccessPath,Add-PhysicalDisk,Add-Printer,Add-PrinterDriver,Add-PrinterPort,Add-StorageFaultDomain,Add-TargetPortToMaskingSet,Add-VirtualDiskToMaskingSet,Add-VpnConnection,Add-VpnConnectionRoute,Add-VpnConnectionTriggerApplication,Add-VpnConnectionTriggerDnsConfiguration,Add-VpnConnectionTriggerTrustedNetwork,AfterAll,AfterEach,Assert-MockCalled,Assert-VerifiableMocks,Backup-BitLockerKeyProtector,BackupToAAD-BitLockerKeyProtector,BeforeAll,BeforeEach,Block-FileShareAccess,Block-SmbShareAccess,Clear-BitLockerAutoUnlock,Clear-Disk,Clear-DnsClientCache,Clear-FileStorageTier,Clear-Host,Clear-PcsvDeviceLog,Clear-StorageDiagnosticInfo,Close-SmbOpenFile,Close-SmbSession,Compress-Archive,Configuration,Connect-IscsiTarget,Connect-VirtualDisk,Context,convert,ConvertFrom-SddlString,Copy-NetFirewallRule,Copy-NetIPsecMainModeCryptoSet,Copy-NetIPsecMainModeRule,Copy-NetIPsecPhase1AuthSet,Copy-NetIPsecPhase2AuthSet,Copy-NetIPsecQuickModeCryptoSet,Copy-NetIPsecRule,Debug-FileShare,Debug-MMAppPrelaunch,Debug-StorageSubSystem,Debug-Volume,Describe,Disable-BitLocker,Disable-BitLockerAutoUnlock,Disable-DAManualEntryPointSelection,Disable-Dsebug,Disable-MMAgent,Disable-NetAdapter,Disable-NetAdapterBinding,Disable-NetAdapterChecksumOffload,Disable-NetAdapterEncapsulatedPacketTaskOffload,Disable-NetAdapterIPsecOffload,Disable-NetAdapterLso,Disable-NetAdapterPacketDirect,Disable-NetAdapterPowerManagement,Disable-NetAdapterQos,Disable-NetAdapterRdma,Disable-NetAdapterRsc,Disable-NetAdapterRss,Disable-NetAdapterSriov,Disable-NetAdapterVmq,Disable-NetDnsTransitionConfiguration,Disable-NetFirewallRule,Disable-NetIPHttpsProfile,Disable-NetIPsecMainModeRule,Disable-NetIPsecRule,Disable-NetNatTransitionConfiguration,Disable-NetworkSwitchEthernetPort,Disable-NetworkSwitchFeature,Disable-NetworkSwitchVlan,Disable-OdbcPerfCounter,Disable-PhysicalDiskIdentification,Disable-PnpDevice,Disable-PSTrace,Disable-PSWSManCombinedTrace,Disable-ScheduledTask,Disable-SmbDelegation,Disable-StorageEnclosureIdentification,Disable-StorageEnclosurePower,Disable-StorageHighAvailability,Disable-StorageMaintenanceMode,Disable-WdacBidTrace,Disable-WSManTrace,Disconnect-IscsiTarget,Disconnect-VirtualDisk,Dismount-DiskImage,Enable-BitLocker,Enable-BitLockerAutoUnlock,Enable-DAManualEntryPointSelection,Enable-Dsebug,Enable-MMAgent,Enable-NetAdapter,Enable-NetAdapterBinding,Enable-NetAdapterChecksumOffload,Enable-NetAdapterEncapsulatedPacketTaskOffload,Enable-NetAdapterIPsecOffload,Enable-NetAdapterLso,Enable-NetAdapterPacketDirect,Enable-NetAdapterPowerManagement,Enable-NetAdapterQos,Enable-NetAdapterRdma,Enable-NetAdapterRsc,Enable-NetAdapterRss,Enable-NetAdapterSriov,Enable-NetAdapterVmq,Enable-NetDnsTransitionConfiguration,Enable-NetFirewallRule,Enable-NetIPHttpsProfile,Enable-NetIPsecMainModeRule,Enable-NetIPsecRule,Enable-NetNatTransitionConfiguration,Enable-NetworkSwitchEthernetPort,Enable-NetworkSwitchFeature,Enable-NetworkSwitchVlan,Enable-OdbcPerfCounter,Enable-PhysicalDiskIdentification,Enable-PnpDevice,Enable-PSTrace,Enable-PSWSManCombinedTrace,Enable-ScheduledTask,Enable-SmbDelegation,Enable-StorageEnclosureIdentification,Enable-StorageEnclosurePower,Enable-StorageHighAvailability,Enable-StorageMaintenanceMode,Enable-WdacBidTrace,Enable-WSManTrace,Expand-Archive,Export-ODataEndpointProxy,Export-ScheduledTask,Find-Command,Find-DscResource,Find-Module,Find-NetIPsecRule,Find-NetRoute,Find-RoleCapability,Find-Script,Flush-EtwTraceSession,Format-Hex,Format-Volume,Get-AppBackgroundTask,Get-AppxLastError,Get-AppxLog,Get-AutologgerConfig,Get-BitLockerVolume,Get-ClusteredScheduledTask,Get-DAClientExperienceConfiguration,Get-DAConnectionStatus,Get-DAEntryPointTableItem,Get-DedupProperties,Get-Disk,Get-DiskImage,Get-DiskStorageNodeView,Get-DnsClient,Get-DnsClientCache,Get-DnsClientGlobalSetting,Get-DnsClientNrptGlobal,Get-DnsClientNrptPolicy,Get-DnsClientNrptRule,Get-DnsClientServerAddress,Get-DscConfiguration,Get-DscConfigurationStatus,Get-DscLocalConfigurationManager,Get-DscResource,Get-Dtc,Get-DtcAdvancedHostSetting,Get-DtcAdvancedSetting,Get-DtcClusterDefault,Get-DtcClusterTMMapping,Get-Dtefault,Get-DtcLog,Get-DtcNetworkSetting,Get-DtcTransaction,Get-DtcTransactionsStatistics,Get-DtcTransactionsTraceSession,Get-DtcTransactionsTraceSetting,Get-EtwTraceProvider,Get-EtwTraceSession,Get-FileHash,Get-FileIntegrity,Get-FileShare,Get-FileShareAccessControlEntry,Get-FileStorageTier,Get-InitiatorId,Get-InitiatorPort,Get-InstalledModule,Get-InstalledScript,Get-IscsiConnection,Get-IscsiSession,Get-IscsiTarget,Get-IscsiTargetPortal,Get-IseSnippet,Get-LogProperties,Get-MaskingSet,Get-MMAgent,Get-MockDynamicParameters,Get-MpComputerStatus,Get-MpPreference,Get-MpThreat,Get-MpThreatCatalog,Get-MpThreatDetection,Get-NCSIPolicyConfiguration,Get-Net6to4Configuration,Get-NetAdapter,Get-NetAdapterAdvancedProperty,Get-NetAdapterBinding,Get-NetAdapterChecksumOffload,Get-NetAdapterEncapsulatedPacketTaskOffload,Get-NetAdapterHardwareInfo,Get-NetAdapterIPsecOffload,Get-NetAdapterLso,Get-NetAdapterPacketDirect,Get-NetAdapterPowerManagement,Get-NetAdapterQos,Get-NetAdapterRdma,Get-NetAdapterRsc,Get-NetAdapterRss,Get-NetAdapterSriov,Get-NetAdapterSriovVf,Get-NetAdapterStatistics,Get-NetAdapterVmq,Get-NetAdapterVMQQueue,Get-NetAdapterVPort,Get-NetCompartment,Get-NetConnectionProfile,Get-NetDnsTransitionConfiguration,Get-NetDnsTransitionMonitoring,Get-NetEventNetworkAdapter,Get-NetEventPacketCaptureProvider,Get-NetEventProvider,Get-NetEventSession,Get-NetEventVFPProvider,Get-NetEventVmNetworkAdapter,Get-NetEventVmSwitch,Get-NetEventVmSwitchProvider,Get-NetEventWFPCaptureProvider,Get-NetFirewallAddressFilter,Get-NetFirewallApplicationFilter,Get-NetFirewallInterfaceFilter,Get-NetFirewallInterfaceTypeFilter,Get-NetFirewallPortFilter,Get-NetFirewallProfile,Get-NetFirewallRule,Get-NetFirewallSecurityFilter,Get-NetFirewallServiceFilter,Get-NetFirewallSetting,Get-NetIPAddress,Get-NetIPConfiguration,Get-NetIPHttpsConfiguration,Get-NetIPHttpsState,Get-NetIPInterface,Get-NetIPseospSetting,Get-NetIPsecMainModeCryptoSet,Get-NetIPsecMainModeRule,Get-NetIPsecMainModeSA,Get-NetIPsecPhase1AuthSet,Get-NetIPsecPhase2AuthSet,Get-NetIPsecQuickModeCryptoSet,Get-NetIPsecQuickModeSA,Get-NetIPsecRule,Get-NetIPv4Protocol,Get-NetIPv6Protocol,Get-NetIsatapConfiguration,Get-NetLbfoTeam,Get-NetLbfoTeamMember,Get-NetLbfoTeamNic,Get-NetNat,Get-NetNatExternalAddress,Get-NetNatGlobal,Get-NetNatSession,Get-NetNatStaticMapping,Get-NetNatTransitionConfiguration,Get-NetNatTransitionMonitoring,Get-NetNeighbor,Get-NetOffloadGlobalSetting,Get-NetPrefixPolicy,Get-NetQosPolicy,Get-NetRoute,Get-NetSwitchTeam,Get-NetSwitchTeamMember,Get-NetTCPConnection,Get-NetTCPSetting,Get-NetTeredoConfiguration,Get-NetTeredoState,Get-NetTransportFilter,Get-NetUDPEndpoint,Get-NetUDPSetting,Get-NetworkSwitchEthernetPort,Get-NetworkSwitchFeature,Get-NetworkSwitchGlobalData,Get-NetworkSwitchVlan,Get-Odbriver,Get-Odbsn,Get-OdbcPerfCounter,Get-OffloadDataTransferSetting,Get-OperationValidation,Get-Partition,Get-PartitionSupportedSize,Get-PcsvDevice,Get-PcsvDeviceLog,Get-PhysicalDisk,Get-PhysicalDiskStorageNodeView,Get-PhysicalExtent,Get-PhysicalExtentAssociation,Get-PnpDevice,Get-PnpDeviceProperty,Get-PrintConfiguration,Get-Printer,Get-PrinterDriver,Get-PrinterPort,Get-PrinterProperty,Get-PrintJob,Get-PSRepository,Get-ResiliencySetting,Get-ScheduledTask,Get-ScheduledTaskInfo,Get-SmbBandWidthLimit,Get-SmbClientConfiguration,Get-SmbClientNetworkInterface,Get-SmbConnection,Get-SmbDelegation,Get-SmbGlobalMapping,Get-SmbMapping,Get-SmbMultichannelConnection,Get-SmbMultichannelConstraint,Get-SmbOpenFile,Get-SmbServerConfiguration,Get-SmbServerNetworkInterface,Get-SmbSession,Get-SmbShare,Get-SmbShareAccess,Get-SmbWitnessClient,Get-StartApps,Get-StorageAdvancedProperty,Get-StorageDiagnosticInfo,Get-StorageEnclosure,Get-StorageEnclosureStorageNodeView,Get-StorageEnclosureVendorData,Get-StorageExtendedStatus,Get-StorageFaultDomain,Get-StorageFileServer,Get-StorageFirmwareInformation,Get-StorageHealthAction,Get-StorageHealthReport,Get-StorageHealthSetting,Get-StorageJob,Get-StorageNode,Get-StoragePool,Get-StorageProvider,Get-StorageReliabilityCounter,Get-StorageSetting,Get-StorageSubSystem,Get-StorageTier,Get-StorageTierSupportedSize,Get-SupportedClusterSizes,Get-SupportedFileSystems,Get-TargetPort,Get-TargetPortal,Get-TestDriveItem,Get-Verb,Get-VirtualDisk,Get-VirtualDiskSupportedSize,Get-Volume,Get-VolumeCorruptionCount,Get-VolumeScrubPolicy,Get-VpnConnection,Get-VpnConnectionTrigger,Get-WdacBidTrace,Get-WindowsUpdateLog,Get-WUAVersion,Get-WUIsPendingReboot,Get-WULastInstallationDate,Get-WULastScanSuccessDate,Grant-FileShareAccess,Grant-SmbShareAccess,help,Hide-VirtualDisk,Import-IseSnippet,Import-PowerShellDataFile,ImportSystemModules,In,Initialize-Disk,InModuleScope,Install-Dtc,Install-Module,Install-Script,Install-WUUpdates,Invoke-AsWorkflow,Invoke-Mock,Invoke-OperationValidation,Invoke-Pester,It,Lock-BitLocker,mkdir,Mock,more,Mount-DiskImage,Move-SmbWitnessClient,New-AutologgerConfig,New-DAEntryPointTableItem,New-DscChecksum,New-EapConfiguration,New-EtwTraceSession,New-FileShare,New-Fixture,New-Guid,New-IscsiTargetPortal,New-IseSnippet,New-MaskingSet,New-NetAdapterAdvancedProperty,New-NetEventSession,New-NetFirewallRule,New-NetIPAddress,New-NetIPHttpsConfiguration,New-NetIPseospSetting,New-NetIPsecMainModeCryptoSet,New-NetIPsecMainModeRule,New-NetIPsecPhase1AuthSet,New-NetIPsecPhase2AuthSet,New-NetIPsecQuickModeCryptoSet,New-NetIPsecRule,New-NetLbfoTeam,New-NetNat,New-NetNatTransitionConfiguration,New-NetNeighbor,New-NetQosPolicy,New-NetRoute,New-NetSwitchTeam,New-NetTransportFilter,New-NetworkSwitchVlan,New-Partition,New-PesterOption,New-PSWorkflowSession,New-ScheduledTask,New-ScheduledTaskAction,New-ScheduledTaskPrincipal,New-ScheduledTaskSettingsSet,New-ScheduledTaskTrigger,New-ScriptFileInfo,New-SmbGlobalMapping,New-SmbMapping,New-SmbMultichannelConstraint,New-SmbShare,New-StorageFileServer,New-StoragePool,New-StorageSubsystemVirtualDisk,New-StorageTier,New-TemporaryFile,New-VirtualDisk,New-VirtualDiskClone,New-VirtualDiskSnapshot,New-Volume,New-VpnServerAddress,Open-NetGPO,Optimize-StoragePool,Optimize-Volume,oss,Pause,prompt,PSConsoleHostReadline,Publish-Module,Publish-Script,Read-PrinterNfcTag,Register-ClusteredScheduledTask,Register-DnsClient,Register-IscsiSession,Register-PSRepository,Register-ScheduledTask,Register-StorageSubsystem,Remove-AutologgerConfig,Remove-BitLockerKeyProtector,Remove-DAEntryPointTableItem,Remove-DnsClientNrptRule,Remove-DscConfigurationDocument,Remove-DtcClusterTMMapping,Remove-EtwTraceProvider,Remove-FileShare,Remove-InitiatorId,Remove-InitiatorIdFromMaskingSet,Remove-IscsiTargetPortal,Remove-MaskingSet,Remove-MpPreference,Remove-MpThreat,Remove-NetAdapterAdvancedProperty,Remove-NetEventNetworkAdapter,Remove-NetEventPacketCaptureProvider,Remove-NetEventProvider,Remove-NetEventSession,Remove-NetEventVFPProvider,Remove-NetEventVmNetworkAdapter,Remove-NetEventVmSwitch,Remove-NetEventVmSwitchProvider,Remove-NetEventWFPCaptureProvider,Remove-NetFirewallRule,Remove-NetIPAddress,Remove-NetIPHttpsCertBinding,Remove-NetIPHttpsConfiguration,Remove-NetIPseospSetting,Remove-NetIPsecMainModeCryptoSet,Remove-NetIPsecMainModeRule,Remove-NetIPsecMainModeSA,Remove-NetIPsecPhase1AuthSet,Remove-NetIPsecPhase2AuthSet,Remove-NetIPsecQuickModeCryptoSet,Remove-NetIPsecQuickModeSA,Remove-NetIPsecRule,Remove-NetLbfoTeam,Remove-NetLbfoTeamMember,Remove-NetLbfoTeamNic,Remove-NetNat,Remove-NetNatExternalAddress,Remove-NetNatStaticMapping,Remove-NetNatTransitionConfiguration,Remove-NetNeighbor,Remove-NetQosPolicy,Remove-NetRoute,Remove-NetSwitchTeam,Remove-NetSwitchTeamMember,Remove-NetTransportFilter,Remove-NetworkSwitchEthernetPortIPAddress,Remove-NetworkSwitchVlan,Remove-Odbsn,Remove-Partition,Remove-PartitionAccessPath,Remove-PhysicalDisk,Remove-Printer,Remove-PrinterDriver,Remove-PrinterPort,Remove-PrintJob,Remove-SmbBandwidthLimit,Remove-SmbGlobalMapping,Remove-SmbMapping,Remove-SmbMultichannelConstraint,Remove-SmbShare,Remove-StorageFaultDomain,Remove-StorageFileServer,Remove-StorageHealthIntent,Remove-StorageHealthSetting,Remove-StoragePool,Remove-StorageTier,Remove-TargetPortFromMaskingSet,Remove-VirtualDisk,Remove-VirtualDiskFromMaskingSet,Remove-VpnConnection,Remove-VpnConnectionRoute,Remove-VpnConnectionTriggerApplication,Remove-VpnConnectionTriggerDnsConfiguration,Remove-VpnConnectionTriggerTrustedNetwork,Rename-DAEntryPointTableItem,Rename-MaskingSet,Rename-NetAdapter,Rename-NetFirewallRule,Rename-NetIPHttpsConfiguration,Rename-NetIPsecMainModeCryptoSet,Rename-NetIPsecMainModeRule,Rename-NetIPsecPhase1AuthSet,Rename-NetIPsecPhase2AuthSet,Rename-NetIPsecQuickModeCryptoSet,Rename-NetIPsecRule,Rename-NetLbfoTeam,Rename-NetSwitchTeam,Rename-Printer,Repair-FileIntegrity,Repair-VirtualDisk,Repair-Volume,Reset-DAClientExperienceConfiguration,Reset-DAEntryPointTableItem,Reset-DtcLog,Reset-NCSIPolicyConfiguration,Reset-Net6to4Configuration,Reset-NetAdapterAdvancedProperty,Reset-NetDnsTransitionConfiguration,Reset-NetIPHttpsConfiguration,Reset-NetIsatapConfiguration,Reset-NetTeredoConfiguration,Reset-PhysicalDisk,Reset-StorageReliabilityCounter,Resize-Partition,Resize-StorageTier,Resize-VirtualDisk,Restart-NetAdapter,Restart-PcsvDevice,Restart-PrintJob,Restore-DscConfiguration,Restore-NetworkSwitchConfiguration,Resume-BitLocker,Resume-PrintJob,Revoke-FileShareAccess,Revoke-SmbShareAccess,SafeGetCommand,Save-EtwTraceSession,Save-Module,Save-NetGPO,Save-NetworkSwitchConfiguration,Save-Script,Send-EtwTraceSession,Set-AutologgerConfig,Set-ClusteredScheduledTask,Set-DAClientExperienceConfiguration,Set-DAEntryPointTableItem,Set-Disk,Set-DnsClient,Set-DnsClientGlobalSetting,Set-DnsClientNrptGlobal,Set-DnsClientNrptRule,Set-DnsClientServerAddress,Set-DtcAdvancedHostSetting,Set-DtcAdvancedSetting,Set-DtcClusterDefault,Set-DtcClusterTMMapping,Set-Dtefault,Set-DtcLog,Set-DtcNetworkSetting,Set-DtcTransaction,Set-DtcTransactionsTraceSession,Set-DtcTransactionsTraceSetting,Set-DynamicParameterVariables,Set-EtwTraceProvider,Set-FileIntegrity,Set-FileShare,Set-FileStorageTier,Set-InitiatorPort,Set-IscsiChapSecret,Set-LogProperties,Set-MMAgent,Set-MpPreference,Set-NCSIPolicyConfiguration,Set-Net6to4Configuration,Set-NetAdapter,Set-NetAdapterAdvancedProperty,Set-NetAdapterBinding,Set-NetAdapterChecksumOffload,Set-NetAdapterEncapsulatedPacketTaskOffload,Set-NetAdapterIPsecOffload,Set-NetAdapterLso,Set-NetAdapterPacketDirect,Set-NetAdapterPowerManagement,Set-NetAdapterQos,Set-NetAdapterRdma,Set-NetAdapterRsc,Set-NetAdapterRss,Set-NetAdapterSriov,Set-NetAdapterVmq,Set-NetConnectionProfile,Set-NetDnsTransitionConfiguration,Set-NetEventPacketCaptureProvider,Set-NetEventProvider,Set-NetEventSession,Set-NetEventVFPProvider,Set-NetEventVmSwitchProvider,Set-NetEventWFPCaptureProvider,Set-NetFirewallAddressFilter,Set-NetFirewallApplicationFilter,Set-NetFirewallInterfaceFilter,Set-NetFirewallInterfaceTypeFilter,Set-NetFirewallPortFilter,Set-NetFirewallProfile,Set-NetFirewallRule,Set-NetFirewallSecurityFilter,Set-NetFirewallServiceFilter,Set-NetFirewallSetting,Set-NetIPAddress,Set-NetIPHttpsConfiguration,Set-NetIPInterface,Set-NetIPseospSetting,Set-NetIPsecMainModeCryptoSet,Set-NetIPsecMainModeRule,Set-NetIPsecPhase1AuthSet,Set-NetIPsecPhase2AuthSet,Set-NetIPsecQuickModeCryptoSet,Set-NetIPsecRule,Set-NetIPv4Protocol,Set-NetIPv6Protocol,Set-NetIsatapConfiguration,Set-NetLbfoTeam,Set-NetLbfoTeamMember,Set-NetLbfoTeamNic,Set-NetNat,Set-NetNatGlobal,Set-NetNatTransitionConfiguration,Set-NetNeighbor,Set-NetOffloadGlobalSetting,Set-NetQosPolicy,Set-NetRoute,Set-NetTCPSetting,Set-NetTeredoConfiguration,Set-NetUDPSetting,Set-NetworkSwitchEthernetPortIPAddress,Set-NetworkSwitchPortMode,Set-NetworkSwitchPortProperty,Set-NetworkSwitchVlanProperty,Set-Odbriver,Set-Odbsn,Set-Partition,Set-PcsvDeviceBootConfiguration,Set-PcsvDeviceNetworkConfiguration,Set-PcsvDeviceUserPassword,Set-PhysicalDisk,Set-PrintConfiguration,Set-Printer,Set-PrinterProperty,Set-PSRepository,Set-ResiliencySetting,Set-ScheduledTask,Set-SmbBandwidthLimit,Set-SmbClientConfiguration,Set-SmbPathAcl,Set-SmbServerConfiguration,Set-SmbShare,Set-StorageFileServer,Set-StorageHealthSetting,Set-StoragePool,Set-StorageProvider,Set-StorageSetting,Set-StorageSubSystem,Set-StorageTier,Set-TestInconclusive,Setup,Set-VirtualDisk,Set-Volume,Set-VolumeScrubPolicy,Set-VpnConnection,Set-VpnConnectionIPsecConfiguration,Set-VpnConnectionProxy,Set-VpnConnectionTriggerDnsConfiguration,Set-VpnConnectionTriggerTrustedNetwork,Should,Show-NetFirewallRule,Show-NetIPsecRule,Show-VirtualDisk,Start-AppBackgroundTask,Start-AutologgerConfig,Start-Dtc,Start-DtcTransactionsTraceSession,Start-EtwTraceSession,Start-MpScan,Start-MpWDOScan,Start-NetEventSession,Start-PcsvDevice,Start-ScheduledTask,Start-StorageDiagnosticLog,Start-Trace,Start-WUScan,Stop-DscConfiguration,Stop-Dtc,Stop-DtcTransactionsTraceSession,Stop-EtwTraceSession,Stop-NetEventSession,Stop-PcsvDevice,Stop-ScheduledTask,Stop-StorageDiagnosticLog,Stop-StorageJob,Stop-Trace,Suspend-BitLocker,Suspend-PrintJob,Sync-NetIPsecRule,TabExpansion2,Test-Dtc,Test-NetConnection,Test-ScriptFileInfo,Unblock-FileShareAccess,Unblock-SmbShareAccess,Uninstall-Dtc,Uninstall-Module,Uninstall-Script,Unlock-BitLocker,Unregister-AppBackgroundTask,Unregister-ClusteredScheduledTask,Unregister-IscsiSession,Unregister-PSRepository,Unregister-ScheduledTask,Unregister-StorageSubsystem,Update-Disk,Update-DscConfiguration,Update-EtwTraceSession,Update-HostStorageCache,Update-IscsiTarget,Update-IscsiTargetPortal,Update-Module,Update-ModuleManifest,Update-MpSignature,Update-NetIPsecRule,Update-Script,Update-ScriptFileInfo,Update-SmbMultichannelConnection,Update-StorageFirmware,Update-StoragePool,Update-StorageProviderCache,Write-DtcTransactionsTraceSession,Write-PrinterNfcTag,Write-VolumeCache
	},
	morekeywords={Do,Else,For,ForEach,Function,If,In,Until,While},
	alsodigit={-},
	sensitive=false,
	morecomment=[l]{\#},
	morecomment=[n]{<\#}{\#>},
	morestring=[b]{"},
	morestring=[b]{'},
	morestring=[s]{@'}{'@},
	morestring=[s]{@"}{"@}
}
\usepackage{amssymb}
\usepackage{pifont}
\newcommand{\xmark}{\ding{55}}%
\usepackage{graphicx}
\providecommand{\keywords}[1]{\textbf{\textit{Index terms---}} #1}

\begin{document}
\author{Vasilios Koutsokostas$^{1}$ and Constantinos Patsakis$^{1,2}$\\
        \small $^{1}$Department of Informatics, University of Piraeus,\\\small 80 Karaoli \& Dimitriou str., 18534 Piraeus, Greece\\
        \small $^{2}$Information Management Systems Institute,\\\small Athena Research Center, Artemidos 6, Marousi 15125, Greece\\
}
\date{}

\title{Python and Malware: Developing Stealth and Evasive Malware Without Obfuscation}

\maketitle
  \begin{abstract}
    With the continuous rise of malicious campaigns and the exploitation of new attack vectors, it is necessary to assess the efficacy of the defensive mechanisms used to detect them. To this end, the contribution of our work is twofold. First, it introduces a new method for obfuscating malicious code to bypass all static checks of multi-engine scanners, such as VirusTotal. Interestingly, our approach to generating the malicious executables is not based on introducing a new packer but on the augmentation of the capabilities of an existing and widely used tool for packaging Python, PyInstaller but can be used for all similar packaging tools. As we prove, the problem is deeper and inherent in almost all antivirus engines and not PyInstaller specific. Second, our work exposes significant issues of well-known sandboxes that allow malware to evade their checks. As a result, we show that stealth and evasive malware can be efficiently developed, bypassing with ease state of the art malware detection tools without raising any alert.
\end{abstract}

\section{Introduction}
\label{sec:introduction}

Adversaries are continually trying to attack systems, to gain access to information and other resources. This leads to a continuous arms race that has significantly augmented the sophistication of the methods used to penetrate systems and, to a lesser extent, those deployed to protect them. Therefore, novel security mechanisms are developed using advanced methods to detect malicious patterns exploiting all possible features, using machine learning and artificial intelligence methods in the past few years. However, the adversaries are crafting complex new attacks, exploiting the human factor, and often resort to encryption and other obfuscation methods to hide their malicious traffic and actions.

Malware, a piece of software that is crafted to perform a malicious task in a computing system, is a problem which has plagued computing systems for decades. Furthermore, the relatively recent introduction of cryptocurrencies has significantly changed the cybercrime ecosystem as it has provided a simple monetisation method with some privacy guarantees. Indeed, as reported by several sources, cybercrime has become a multi-billion underground economy with such economic impact \cite{crime_report,thomas2020cybercrime} that the World Economic Forum considers it the second most-concerning threat to global commerce over the next decade \cite{wef}.

The main pillars for detecting and analysing malware are static and dynamic analysis \cite{gandotra2014malware}. In the former, there is no execution of the file under inspection. Therefore, we try to correlate patterns in every aspect of the file that can be collected without executing it, including but not limited to imported libraries, file segments, API calls, strings, file structure, entropy, etc. On the contrary, in dynamic analysis, we open and/or execute the file in a controlled testing environment (sandbox) to identify what this piece of software does. In this regard, we keep track of every possible network interaction, filesystem change, memory dump, processes etc. \cite{or2019dynamic}, replicating a real-world host environment. Moreover, one may debug the binary to execute it, possibly line by line, to understand what it does and how, and even manipulate its behaviour.

The above methods are well-known to malware authors who try to bypass them by introducing obfuscation and other anti-analysis methods \cite{branco2012scientific}. Modern malware frequently uses packers and encryption to obfuscate their contents and bypass static analysis checks by generating new binaries with different static properties. Similarly, they are often armoured with evasion methods to bypass dynamic analysis. Thus, they perform specific checks in the system to determine whether they are being executed in a virtual environment and if known protection mechanisms typical of sandboxes are running, and can assess their execution mode to tell whether they are being debugged \cite{issa2012anti}. If any of these checks is positive, the malware typically changes its behaviour to harden and slow down its analysis. All the above can come under one umbrella to facilitate malware evasion by simultaneously packing the binary and armouring it with a myriad of evasion methods \cite{lictua2018anti}.

The main goal of this work is to assess the effort and methods needed to create \textit{stealth malware}. We define this stealth concept in an objective and repeatable way. More precisely, we consider that a malware sample is stealth if (i) it achieves a ``clean sheet'' after inspection by multi-engine scanners, such as VirusTotal (VT) and (ii) malware sandbox environments do not consider it malicious per se.
VT and other similar services used \textit{statically} examine the file with several dozens of antiviruses (AVs). Therefore, even if an AV may detect the malware on execution, VT's verdict might classify it as benign.  Note that a clean sheet verdict from VT, which has around 70 AVs clearly shows the trend of the market, meaning that the rest of the AVs, which are minor share of the market are not expected to have different behaviour. Practically, our work starts from understanding why some AVs are erroneously flagging some executables as malicious and uncovers an inherent problem of AV engines when handling Python files. This can be easily escalated to develop undetectable malware. What is even more alarming is the fact that while one may argue that there are several tricks to bypass static AV tests by hiding the payload, we illustrate that a threat actor does not need to cover the payload. Widely used payloads can be simply embedded in Python and escape the detection.


Nevertheless, it is clear that once the user is lured to execute malware, it might be too late to block its actions. Moreover, we consider the malware as stealth if it escapes detection from the state of the art malware sandboxes. To this end, we experimented with the most well-known sandboxes publicly available on the Internet. Our analysis and experiments have uncovered significant issues in these sandbox environments that allow malware to bypass them. Based on the above, our work illustrates critical issues in detecting malware that affects the whole ecosystem, spanning from how AVs statically recognise malware, to the evasion from sandboxed environments. Practically, using our methods, one may efficiently develop malware or armour an existing one so that that it is not detected by a wide range of state of the art tools used for detecting malware.

In what follows, we provide a brief overview of the related work. Then, we proceed with discussing the conceptual approach for the development of stealth malware. In Section \ref{sec:experiments}, we analyse our experiments and the extracted results. Then, in Section \ref{sec:discussion}, we discuss our findings and their impact. Finally, the article concludes summarising the contributions of our work and streamlining future work.

\noindent\textbf{Ethical compliance:} Our work complies with the standards for conducting offensive security in an ethical way. To this end, we have responsibly disclosed our findings to each sandbox provider individually prior to submitting this work. Moreover, we have not published nor communicated our methods to prevent them from being used in the wild.
\section{Related work}
\label{sec:related}

Similar to the use of sandboxes for cats, a malware sandbox is a controlled virtualised environment in which a potentially dangerous file is submitted for inspection, so that it does not ``litter'' the rest of the system. This environment will automatically execute/open the file and analyse its behaviour, such as filesystem interaction, network connections, registry changes and access, API calls, memory access, etc. The virtualised and isolated nature of the environment prevents the malware from causing any harm to the system performing the analysis. Another approach would be to actually debug the suspicious file and examine in detail command by command and even alter its behaviour.

\begin{figure*}[th]
    \centering
    \includegraphics[width=.8\textwidth]{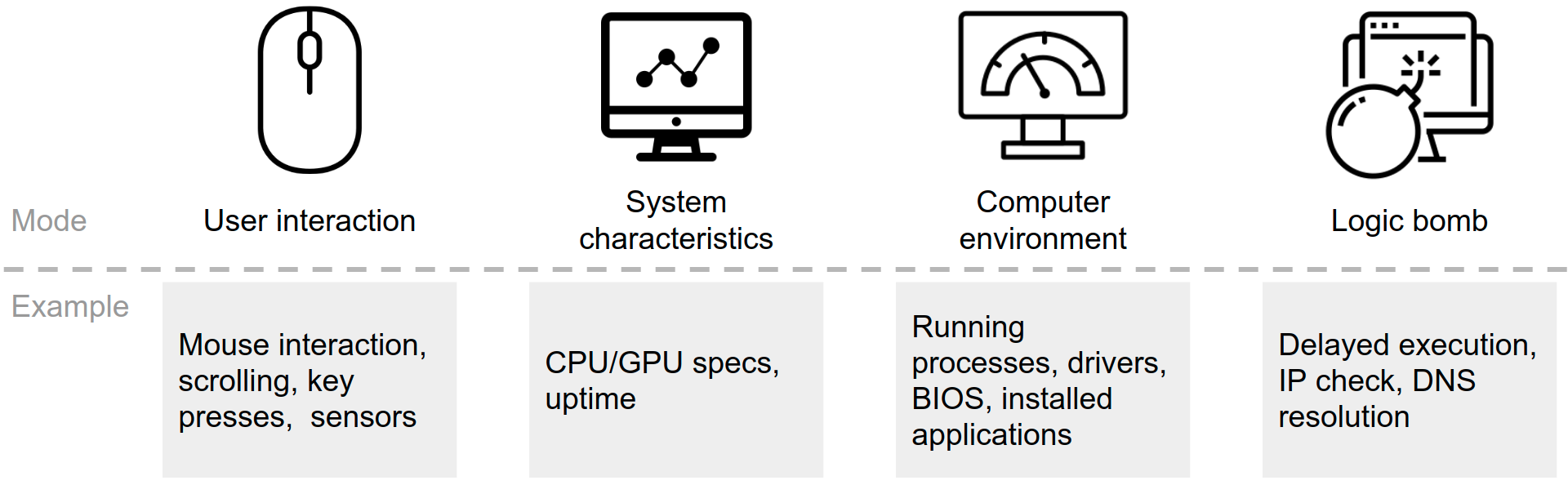}
    \caption{Sandbox evasion methods overview.}
    \label{fig:evasion}
\end{figure*}

Clearly, the above is not the ideal for the adversary, so almost all modern malware come equipped with an evasion method leveraging, for instance, sandbox and debugger detection methods. For the sandbox evasion, the malware performs a broad range of checks to assess the environment they are being executed. In essence, the malware will look for \textit{environmental artifacts} \cite{bulazel2017survey} which include but are not limited to hardware identifiers, presence of user interaction, sensor readings, uptime, usernames, timing discrepancies, registry values, and hardware specifications \cite{martignoni2009testing,shi2014cardinal}, see Figure \ref{fig:evasion}. Therefore, such a malware would resolve to i) calls to the registry, check the process list and filesystem to perform pattern matching against a predefined set of strings ii) time measurements to determine whether the elapsed time is aligned with the expected processing time and iii) detect possible deviations from the outcome of specific commands. The above indicates that minor details, for instance, the MAC address of the network may easily reveal the virtualised environment as well as the list of running processes or inconsistencies in CPU/GPU specifications. Some malware may also use logical bombs to deliver their payload. For instance, the execution can be delayed based on time constraints or enabled only after proper packet receipt from a specific domain. In fact the time that a honeypot devotes for execution of a sample introduces many differences on what data is collected. As recently reported by K{\"u}chler et al. \cite{kuchler2021does}, the bulk most of the malware behavior is observed during the first two minutes of execution, while further actions may take up to ten minutes.

It must be highlighted at this point that due to the monetisation model (discussed later on), a sandbox will not execute and inspect a binary for an arbitrary amount of time. Additionally, to analyse as many samples as possible, it cannot provide all the available system resources. Therefore, by delaying the execution, allocating a lot of space and memory, a malware may evade detection. Thus, the sharing of the processing resources may easily expose the virtualised environment as the VM could report the host's processor with a fragment of the available cores. Recently Huang et al. \cite{huang2020pidicators} introduced PiDicators which do not use API calls but pure assembly code and far fewer checks to determine whether a binary is being executed in a VM triggering far fewer alerts. It has to be noted that the wide adoption of virtualised environments in, e.g. cloud computing, some malware is even more targeted, trying to detect sandboxed environments and not simply virtualised \cite{yokoyama2016sandprint}. For more on evasion methods the interested reader may refer to \cite{chen2008towards,issa2012anti,petsas2014rage,uitto2017survey,veerappan2018taxonomy,afianian2018malware,checkpoint,ApostolopoulosK21}.

These countermeasures from the malware have resulted in the introduction of anti-evasion methods. For instance,
\textsc{MalGene} \cite{kirat2015malgene} performs data flow analysis and data mining on the system calls to determine whether the inspected binary actions could be a result of an evasion method.

VM Cloak \cite{Shi:2017:HAM:3171591.3139292} checks the environment for misconfigurations and differences in execution environments that could reveal that the execution is done in a VM, while Leguesse et al. \cite{leguesse2017androneo} harden Android sandboxes which have more sensors to cover. A widely used project for hiding Windows VMs is A. Ortega's {pafish}\footnote{\url{https://github.com/a0rtega/pafish}} which focuses on the checks that are performed by malware.

Recently, D'Elia et al. \cite{cavalaro} introduced a dynamic binary instrumentation based method, called BluePill which allows analysts to instrument the binaries they are dissecting evasive malware in a stealth way so that they cannot determine that they are being debugged. Nevertheless, this is another part of the continuous battle, bringing, for instance, anti-anti evasion methods in this fight \cite{d2019sok}.

Finally, it should be noted that bare-metal malware execution environments, so the execution is performed in an actual and not virtualised environment, so there is no VM nor sandbox stain to cover, are also considered in the literature \cite{kirat2011barebox,guan2017supporting,kirat2014barecloud,mutti2015baredroid,220231}, nevertheless, they cannot be considered a practical solution for assessing malware samples at the desired rate as they cannot scale efficiently.

\section{Python and PyInstaller}
\label{sec:python}
Python is an interpreted programming language with continuous increasing popularity. Despite its readability and simplicity, it has accumulated several features over the years, making it very attractive for scripting and Rapid Application Development. Currently, it is widely used for server-side web development, machine learning, system scripting and secure software-related engineering, especially offensive.

The fact that Python can be used in all major platforms, as well as the fact that it is easy to write and many exploits and offensive security tools, have been written in Python has pushed a lot of malware authors to write their malware in this programming language\footnote{\url{https://unit42.paloaltonetworks.com/unit-42-technical-analysis-seaduke/},\url{https://blog.talosintelligence.com/2020/04/poetrat-covid-19-lures.html},\url{https://blog.netlab.360.com/not-really-new-pyhton-ddos-bot-n3cr0m0rph-necromorph/},\url{https://www.crowdstrike.com/blog/bears-midst-intrusion-democratic-national-committee/}}. However, we argue that there is another more important issue with Python that makes it more attractive for malware authors. AVs have not properly integrated this attack vector in their scope, as we will show in the next paragraphs.

While Python is preinstalled by default in most Unix-like operating systems, it is not the case of Windows. Moreover, Python, as an interpreted language, does not compile to create an executable. To create an executable from a Python script, there are several options, with the most popular one being PyInstaller. PyInstaller takes as input a Python and tries to discover all its module and library dependencies that are needed to properly execute it. To do this, PyInstaller is recursively looking for imports of the necessary files, until it reaches native Python modules and libraries. Once the dependencies are identified, instead of keeping the Python scripts, PyInstaller keeps the compiled Python scripts (.pyc files), usually referred to as Python bytecode. These files, along with an active Python interpreter and environment in the form of what is called the bootloader, are copied in a folder. Thus, PyInstaller allows the packaging of applications in folders and unique executable files without the need to have Python preinstalled.

The bootloader is the core component of PyInstaller as it prepares the environment for executing the Python code and actually executes it. The bootloader is different for each architecture and highly customizable. Once someone launches a bundled Python application, the bootloader is initiated and spawns another child process of itself. The parent bootloader process handles the signals for the two processes and uncompresses all the .pyc files in a folder named \_MEIxxxxxx in the temp folder of the host, where xxxxxx is a random number. The child process loads the temporary Python environment with all the needed modules and libraries for the script can be imported and executes the script. Once the child process terminates, the parent process will cleanup and terminate as well.

To compress the files and create a single executable, PyInstaller uses two compression methods, ZlibArchives for Python compiled files (executable Python zip archives) and CArchive for all other files. In this work, we delibrately study PyInstaller as beyond being the most widely used solutions for creating executables from Python, many other installers are based on it. Therefore, the issues reported in this case can be escalated to other installers.

\section{Conceptual approach}
Our work's conceptual approach is to progressively determine what triggers detection of a malicious binary in static and dynamic analysis and create patches to remove it. We argue that if VirusTotal and other similar engines consider a binary as benign and the dynamic analysis from a sandbox does not trigger an alert, the binary is deemed benign, even by security savvies. In this regard, a \textit{suspicious} indication of sandbox would be considered simply suspicious. Therefore, it will fall below the detection radars and would be executed by a typical user. While we understand that an anti-malware mechanism may detect it upon execution, this is clearly too late in most cases.

Two individual streams emerged from this basic concept, targeting towards evading each analysis. Once we developed the measures that bypassed each one of them individually, we merged them into a unique binary. Therefore, we will present the approach and experiments individually. As we will detail in the next section, for the static analysis, we uploaded our samples to VirusTotal and used the detection output and classification of each antivirus, the reported YARA rules, as well as the community comments to determine which static properties are the ones that lead to the detection of the malware. To further validate our results, we submitted our results to two more similar engines. For the dynamic analysis with sandboxes, we initially submitted some binaries that collected data from each sandbox environment and then used this as an input to armour our binary with evasion measures. Notably, as discussed later in the article, we identified several important issues for many of the sandboxes that were responsibly communicated to them.

\subsection{Bypassing Static Analysis}
The methodology behind the technique to bypass the static analysis stems from observations on PyInstaller\footnote{\url{https://www.pyinstaller.org/}} 4.0 binaries. To generate an executable, PyInstaller adds a lot of ``noise'' to the generated binaries, from, e.g. the libraries that are appended, and even if the code is not malicious, many AVs falsely treat the executable as malware. In fact, as reported by the community, in numerous occasions even simple ``Hello world'' Python scripts are flagged as malicious by several AVs as they consider binaries generated by PyInstaller as malicious by default.

The latter exhibits an erroneous policy applied by almost all AVs; at least the ones used in VT, when handling binaries produced by PyInstaller. In practice, none of them understands its output; probably because of its overblown added libraries. Therefore, on the one hand, we have most antivirus for which PyInstaller acts like an efficient \textit{packer}, so one can hide arbitrary code in them. On the other hand, other AVs have understood this capacity and immediately flag the binaries as malicious.

\begin{lstlisting}[caption={A typical Powershell reverse shell.},label={lst:shell}]
powershell -NoP -NonI -W Hidden -Exec Bypass -Command New-Object System.Net.Sockets.TCPClient("10.0.0.1",4242);$stream = $client.GetStream();[byte[]]$bytes = 0..65535|%{0};while(($i = $stream.Read($bytes, 0, $bytes.Length)) -ne 0){;$data = (New-Object -TypeName System.Text.ASCIIEncoding).GetString($bytes,0, $i);$sendback = (iex $data 2>&1 | Out-String );$sendback2  = $sendback + "PS " + (pwd).Path + "> ";$sendbyte = ([text.encoding]::ASCII).GetBytes($sendback2);$stream.Write($sendbyte,0,$sendbyte.Length);$stream.Flush()};$client.Close()
\end{lstlisting}

\begin{figure*}[th]
 \centering
 \begin{subfigure}[t]{.32\textwidth}
   \centering
   \includegraphics[width=\linewidth]{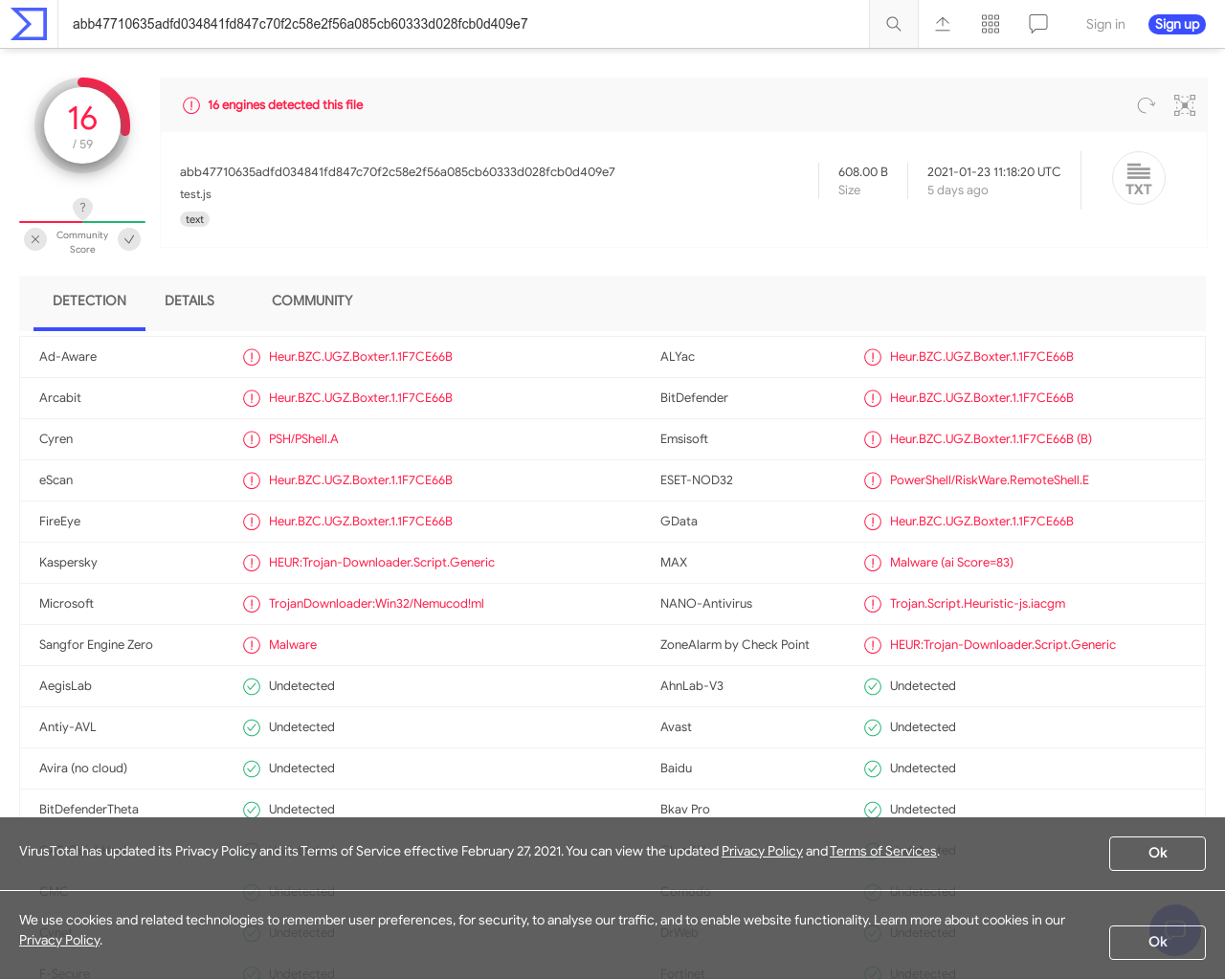}
    \caption{JavaScript using PowerShell reverse shell.
    \protect\url{}}
    \label{fig:jsps}
 \end{subfigure}~
 \begin{subfigure}[t]{.32\textwidth}
   \centering
   \includegraphics[width=\linewidth]{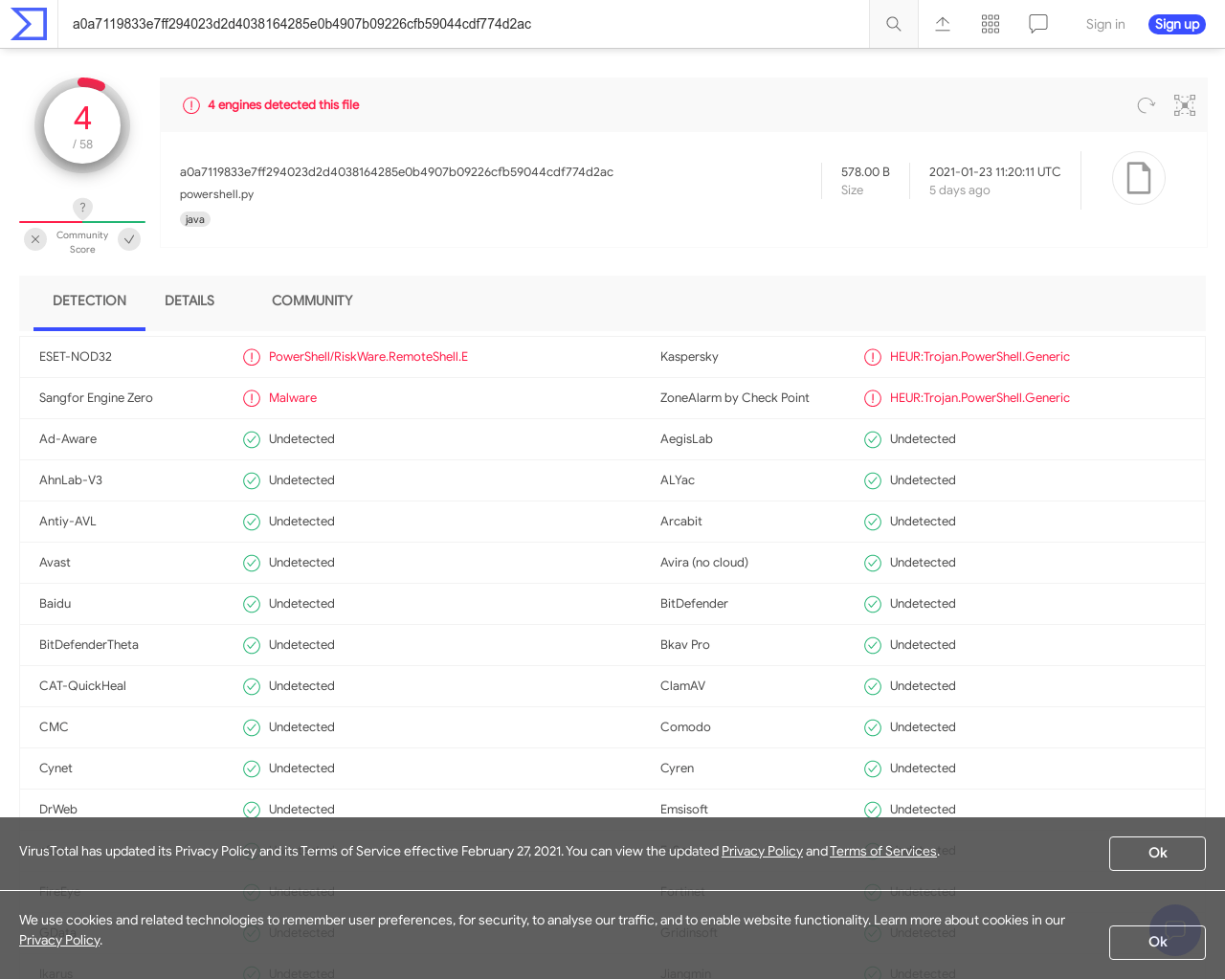}
   \caption{Python using PowerShell reverse shell.
   \protect\url{}}
   \label{fig:pyps}
 \end{subfigure}~
  \begin{subfigure}[t]{.32\textwidth}
   \centering
   \includegraphics[width=\linewidth]{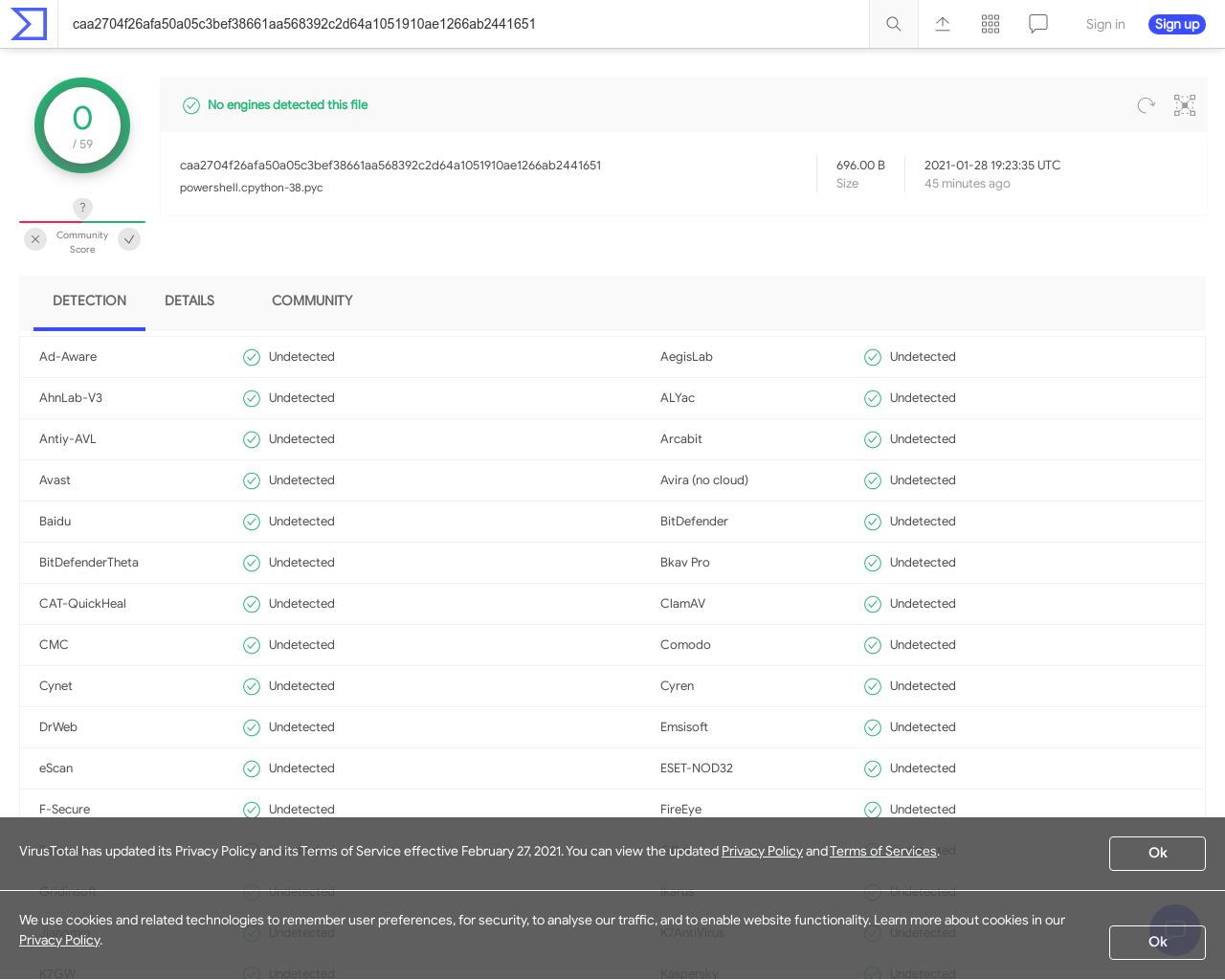}
   \caption{Compiled Python scipt (\texttt{pyc}) of the Python using PowerShell reverse shell script.
   \protect\url{}}
   \label{fig:pycps}
 \end{subfigure}
 \caption{Scan results for reverse shell scripts using Javascript and Python.}
 \label{fig:reverse}
\end{figure*}

In what follows, we dig a bit deeper on the problem with PyInstaller to understand the nature of the noise that makes it act like a packer. We start with a simple reverse shell with a PowerShell script which is typically flagged by AVs. The one-line script is provided in Listing \ref{lst:shell}. Note that similar backdoor mechanisms; e.g. malicious PowerShell execution, are widely used by malware in the wild. Two scripts, one in JavaScript and one in Python were written appending the exact same PowerShell code snippet to their body; therefore, no obfuscation is applied. While both of them are plain ASCII files, with minimal differences in their contents and the malicious string in plain sight, there are significant deviations on their detection from AVs, see Figure \ref{fig:reverse}, which are rather alarming. More precisely, one may observe that the JavaScript file is flagged as malicious by four times more AVs than its Python peer. Notably, none of them were identified correctly, the JavaScript is considered as \textit{text} and the Python as \textit{Java}. While the inconsistency in the detection rate of AVs for almost the same plaintext file cannot be easily understood, the compiled Python file (\texttt{pyc}), and Python bytecode in general, illustrates a more catastrophic result. None of the AVs is able to recognise it as malicious; therefore, it shows that none of the AVs understands what is inside a \texttt{pyc} file as the conversion to the Python compiled file efficiently obfuscates the contents of the script to bypass the static analysis.


The above illustrates a clear strategy to bypass static analysis for an executable. One has to write a Python script which does all the ``dirty job'' and compile it using PyInstaller to hide its malicious content. Then, if we masquerade the PyInstaller enough so that it is not considered as such, we may pass any executable without any detection from the AVs.

Based on the above, our strategy is to exploit these inefficiencies in handling binaries generated by PyInstaller. Thus, the plan is to use PyInstaller to create the binaries out of malicious scripts, but then remove all the possible static features that it appends from the binary. The general outline of the method is illustrated in Algorithm \ref{lbl:staticbypass}.

\subsection{Bypassing Dynamic Analysis}
The dynamic analysis bypass is solely targeted towards bypassing the checks performed by executing the binary in a set of well-known and widely used sandboxes. To this end, we first created a set reconnaissance of executables that were simply collecting environmental data from each sandbox and performing some checks with a standard tool for assessing the sandboxes' quality for malware analysis, pafish\footnote{\url{https://github.com/a0rtega/pafish}}. Once collected, the input was then sent to a server that we controlled to gather and analyse it.

Beyond the output of pafish, which identified several misconfigurations and our own findings, one has to consider some particular inherent issues that such services have. The environmental findings have to be considered in the scope of a service offered in a virtualised environment, for a limited amount of time and with the minimum amount of resources to allow for scaling. As a result, a VM cannot always meet a typical computer's specifications in terms of, e.g. memory, disk, etc.

Finally, one has also to consider that most samples in such a sandbox originate from users without paid plans, so these are tested in VMs that are more limited. Based on the market model (see Section \ref{sec:related}), if a file is considered benign by the static analysis, and the sandboxes have not identified it as malicious, the chances of the file being rescanned in a ``better'' VM drop dramatically.

\section{Experimental results}
\label{sec:experiments}
\subsection{Static Malware Analysis}

\begin{algorithm}[th]
\begin{algorithmic}[1]
\Procedure{Obfuscate\_payload}{\texttt{x}}
  \State Select proper payload\;
  \State Parametrise the payload\;
  \State XOR the payload with a random key\;
  \State Convert the XORed payload to base64\;
\EndProcedure
\Procedure{Patch\_Bootloader}{\texttt{exe}}
  \State Rename PyInstaller references to a random string
  \State Rename files and their calls with pyi\_ prefix to a random prefix.
  \State Replace default icons
  \State Update linker's flags in WScript
\EndProcedure
\Procedure{Patch\_binary}{\texttt{exe}}
  \State Add version to the binary
  \State Remove rich header
  \State Rename \texttt{\_RTDATA} header to \texttt{.bss}
  \State Recalculate PE32 checksum.
\EndProcedure
\State Select payload \texttt{P}
\State \texttt{P'=Obfuscate\_payload(P)}
\State Use a Python \texttt{S} script to call \texttt{P'}\;
\State Build a PE32 executable from \texttt{S} to a single file to generate the bootlader \texttt{B}
\State \texttt{B'=Patch\_Bootloader(B)}
\State Build the PE32 executable \texttt{PE32} from \texttt{S} to a single file with bootlader \texttt{B'}
\State \texttt{PE32'=Patch\_Binary(PE32)}
\end{algorithmic}
\caption{Steps to bypass static analysis}
\label{lbl:staticbypass}
\end{algorithm}
%

Following our findings for the handling of Python bytecode, the main goal of the experiments is to alter the executable in a way that it does not look generated by PyInstaller.
In our experiments, we opted to use some standard malicious payloads as a codebase that were executed through Python, create an executable with the corresponding bootloader of PyInstaller, and then make the necessary changes to the bootloader and the executable to prevent AVs from detecting it.

\begin{figure*}[th]
 \centering
 \begin{subfigure}[t]{.32\textwidth}
   \centering
   \includegraphics[width=\linewidth]{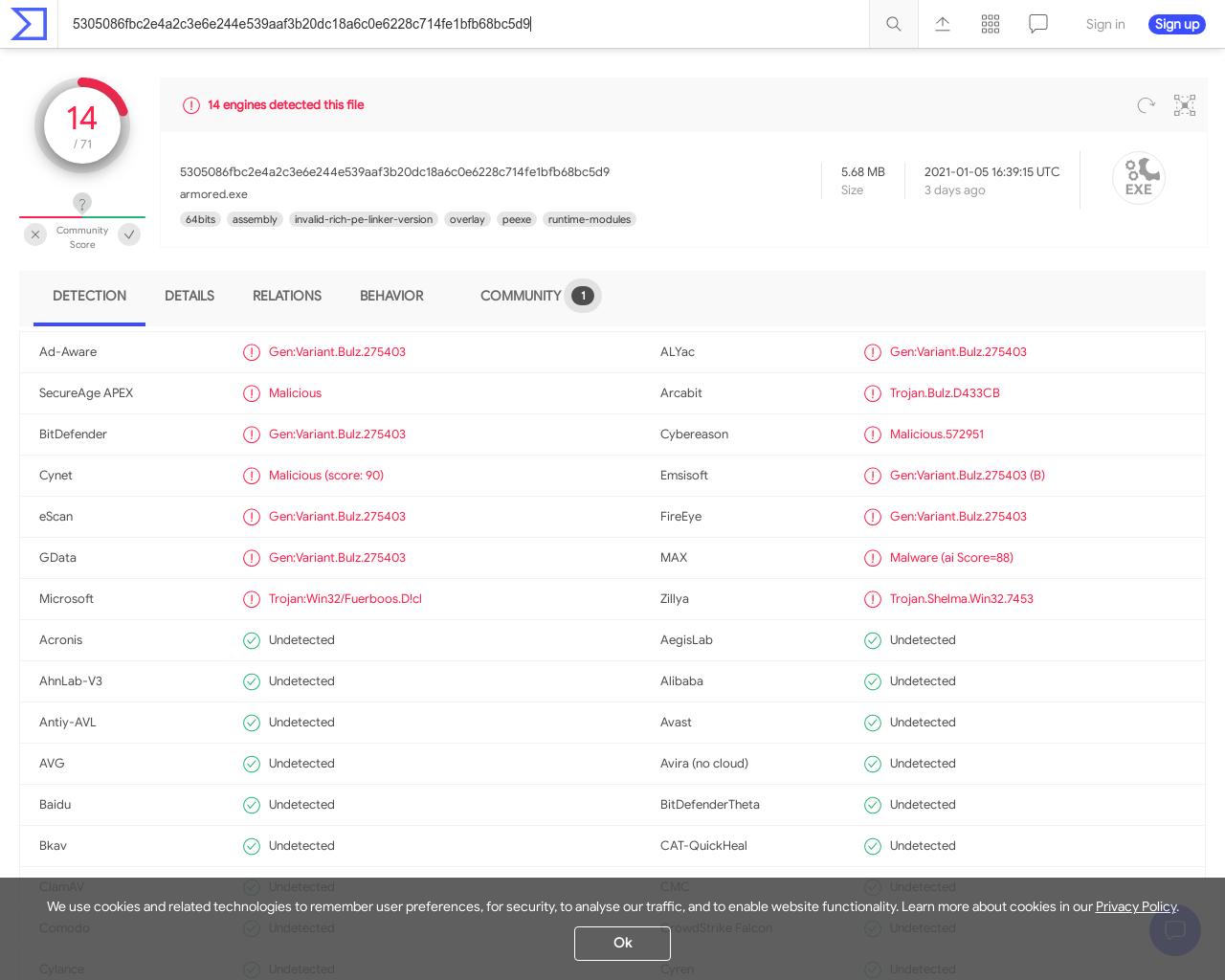}
    \caption{Original binary with PyInstaller}
    \label{fig:sub2}
 \end{subfigure}~
 \begin{subfigure}[t]{.32\textwidth}
   \centering
   \includegraphics[width=\linewidth]{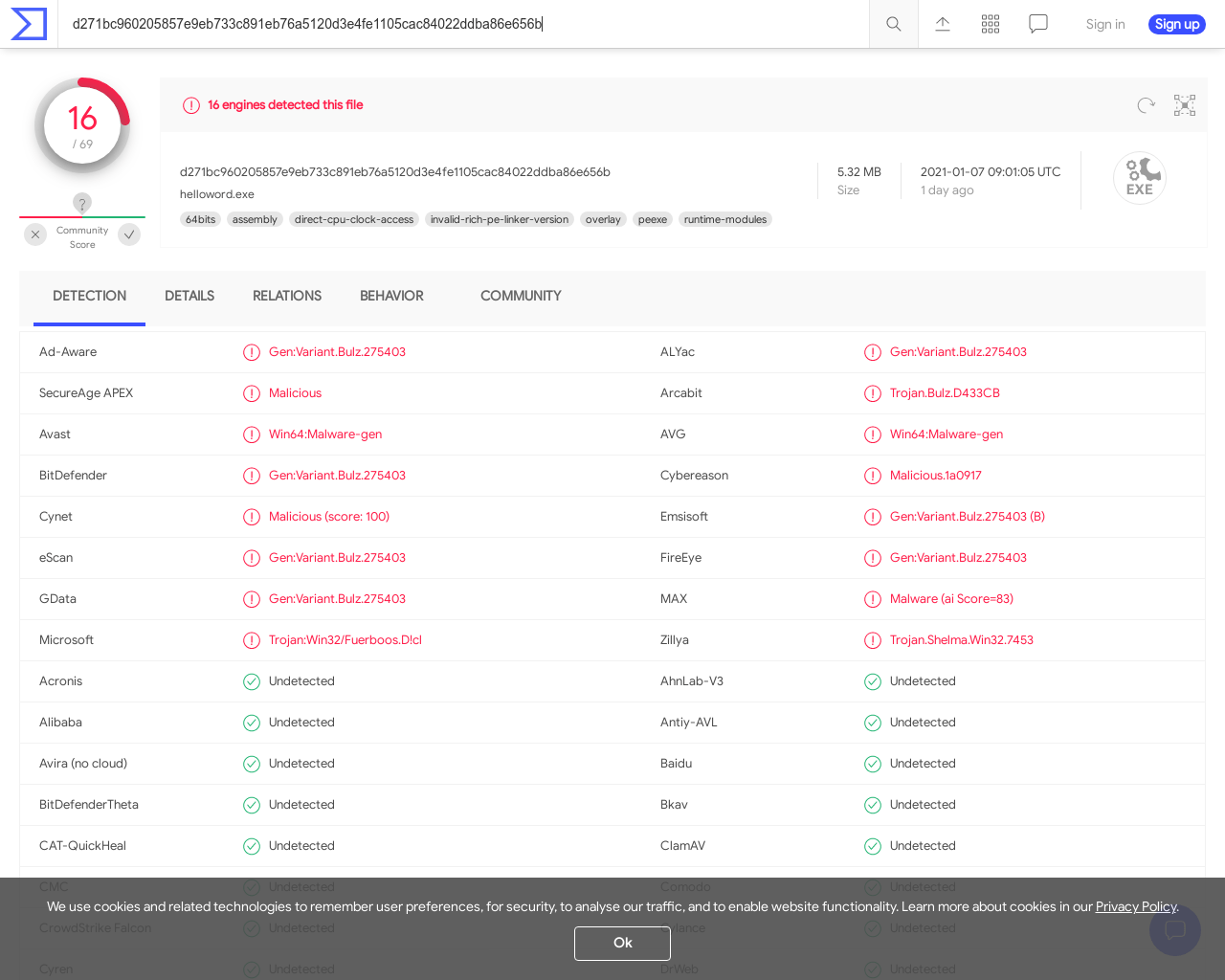}
   \caption{Hello world binary with PyInstaller.}
   \label{fig:sub3}
 \end{subfigure}~
  \begin{subfigure}[t]{.32\textwidth}
   \centering
   \includegraphics[width=\linewidth]{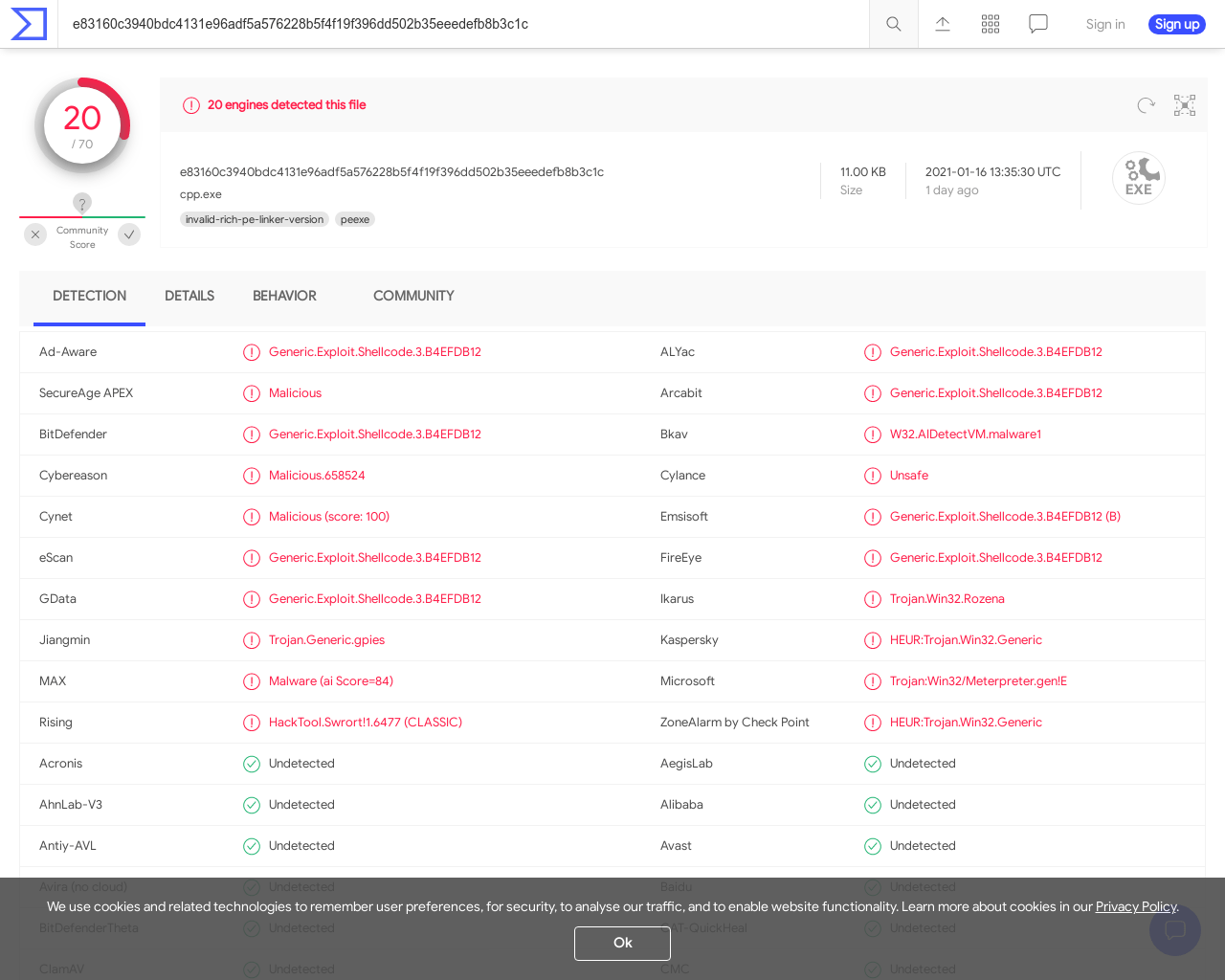}
   \caption{C++ compiled executable with the malicious payload.}
   \label{fig:cpp}
 \end{subfigure}

  \begin{subfigure}[t]{.32\textwidth}
   \centering
   \includegraphics[width=\linewidth]{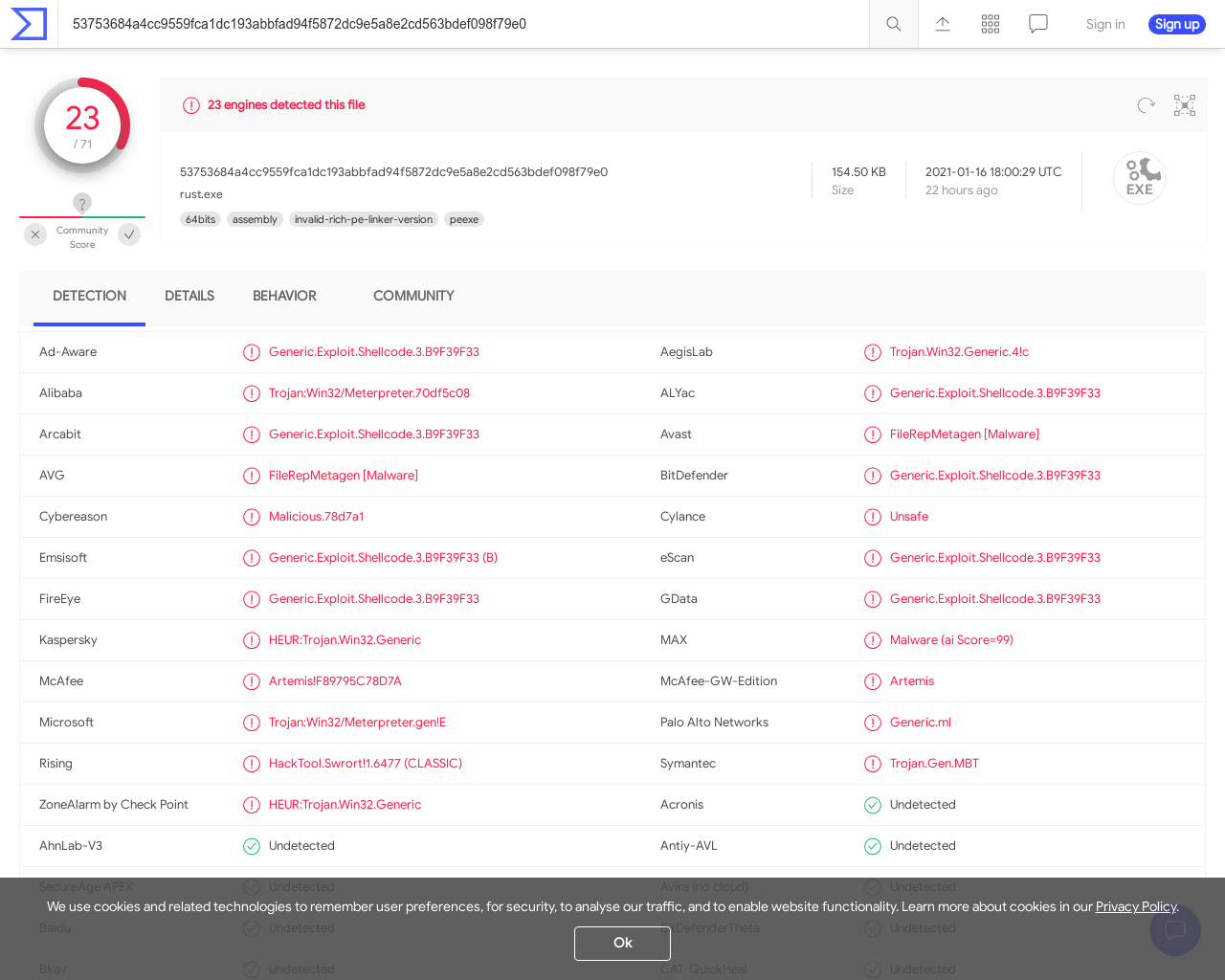}
   \caption{Rust compiled executable with the malicious payload.}
   \label{fig:rust}
 \end{subfigure}~
  \begin{subfigure}[t]{.32\textwidth}
   \centering
   \includegraphics[width=\linewidth]{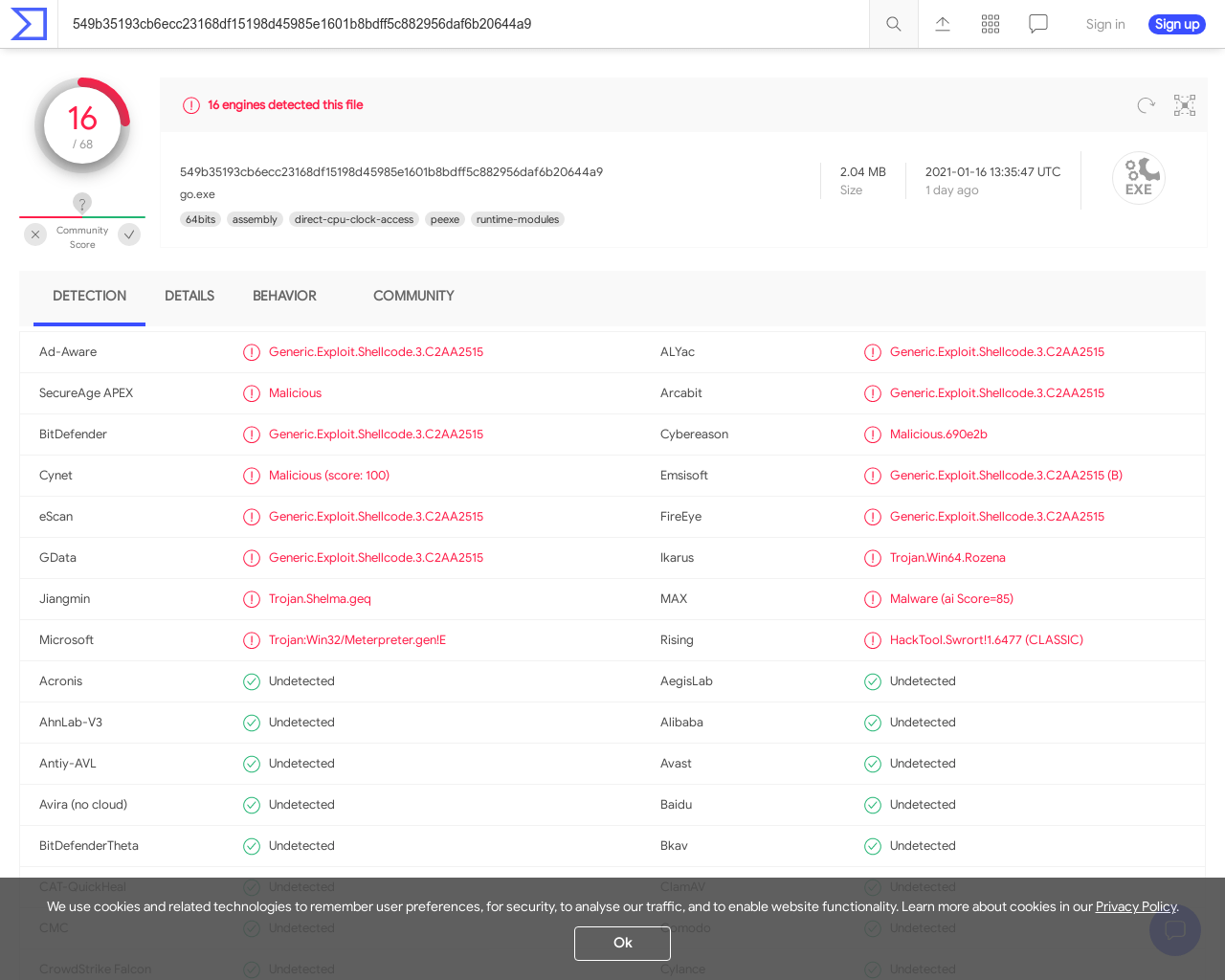}
   \caption{Go compiled executable with the malicious payload.}
   \label{fig:go}
 \end{subfigure}
 \caption{VT detection results for binaries from various languages.}
 \label{fig:initialbinaries}
\end{figure*}

Initially, we wrote a script with a known malicious shellcode payload from \texttt{msfvenom} and a Powershell command that downloads the EICAR anti-malware testfile and XORed that Powershell command with a random hard-coded string and converted it to base64. The reason for these choices is that both of them are well known to trigger AVs; therefore, if any of them is identified by an AV or a sandbox, it will immediately flag the file as malicious in both static and dynamic analysis. We compiled this script with PyInstaller and submitted the executable to VT. As shown in Figure \ref{fig:sub2}, multiple AV engines reported our executable as malicious. Moreover, we scanned a simple ``hello world'' Python script compiled with PyInstaller in VT, and it was also reported as malicious by the same antivirus engines (Figure \ref{fig:sub3}), verifying again the issues described in the previous section. To further validate our results, we created some binaries with the exact same functionality using C++, Rust, and Go and submitted them for analysis to VT, see Figures \ref{fig:cpp}, \ref{fig:rust} and \ref{fig:go} respectively. It is important to highlight in the latter figures that, contrary to the ones for Python, the AVs have correctly identified the presence of shellcode and Meterpreter, as shown by the names that they attribute to our binaries. The difference is rather important since the shellcode is not encoded in any of the implementations showing that PyInstaller has efficiently hidden it from the AVs once again.

\begin{figure}[th]
    \centering
    \includegraphics[width=\linewidth]{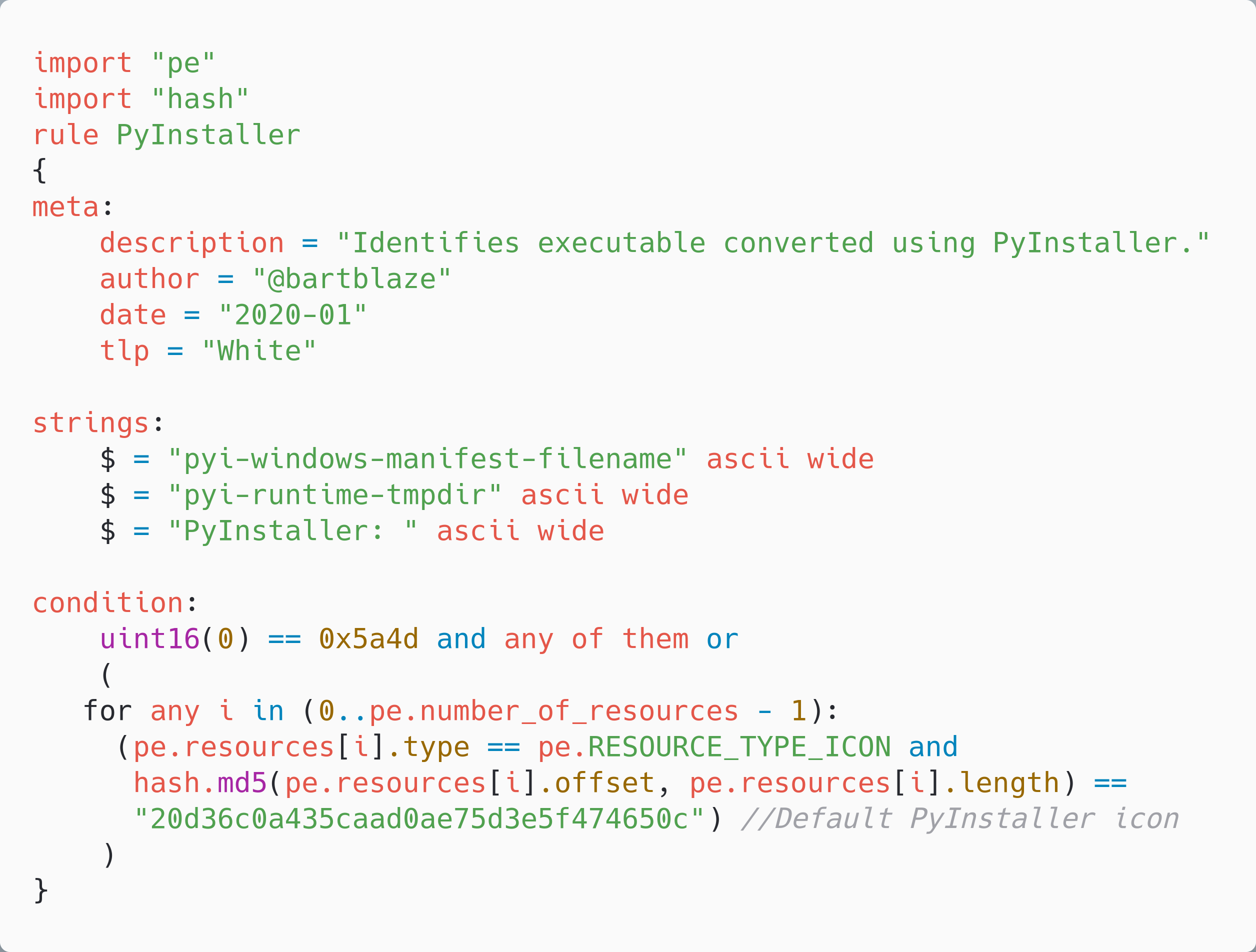}
    \caption{A common YARA rule for detecting PyInstaller. Source: \protect\url{https://github.com/bartblaze/Yara-rules/blob/5f4961049d0d510b11250d5628383398889fc881/rules/generic/PyInstaller.yar}.}
    \label{fig:yararule}
\end{figure}

\begin{figure*}[th]
 \centering
 \begin{subfigure}[t]{.32\textwidth}
   \centering
   \includegraphics[width=\linewidth]{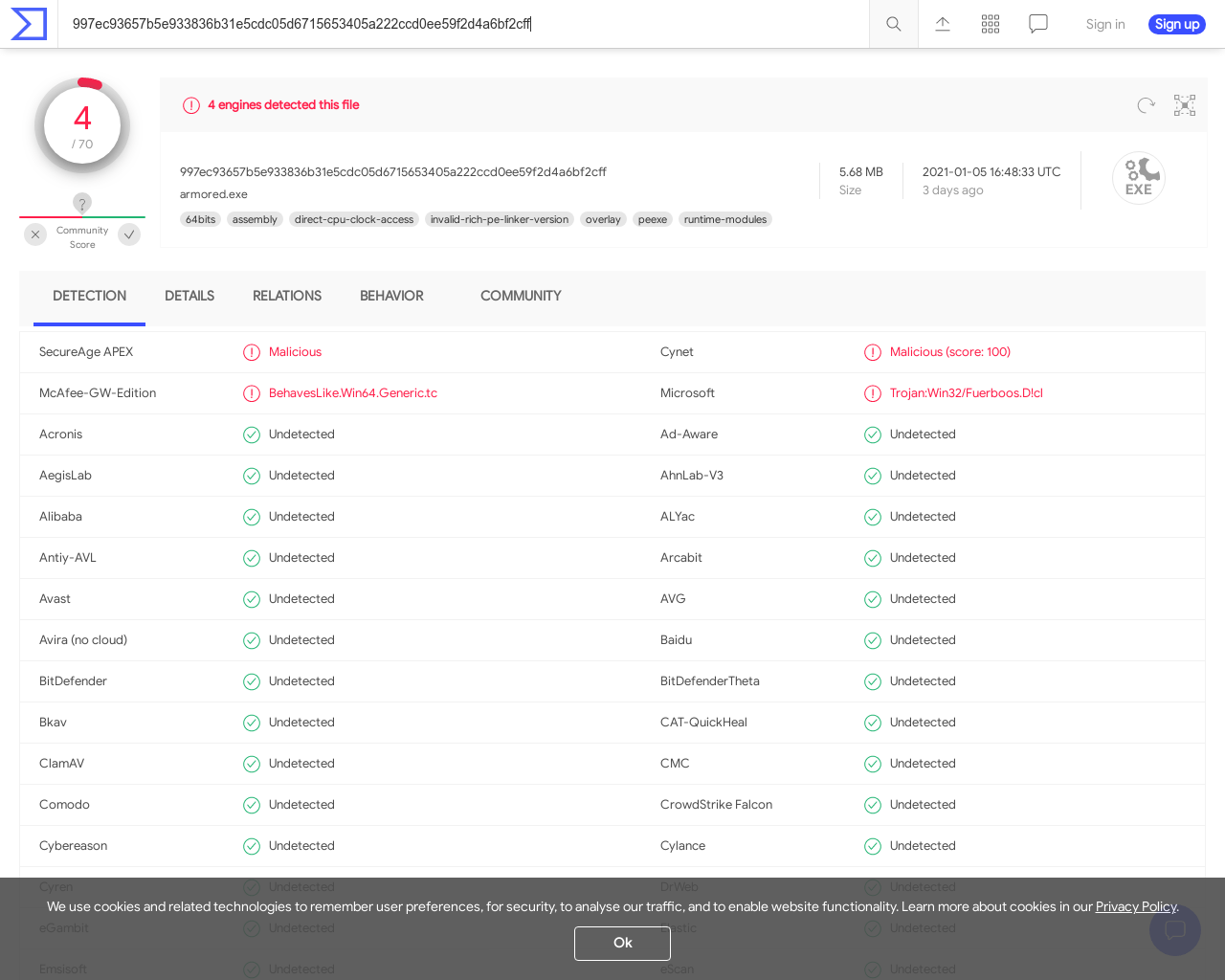}
    \caption{Patched binary with PyInstaller strings removed.}
    \label{fig:sub4}
 \end{subfigure}~
 \begin{subfigure}[t]{.32\textwidth}
   \centering
   \includegraphics[width=\linewidth]{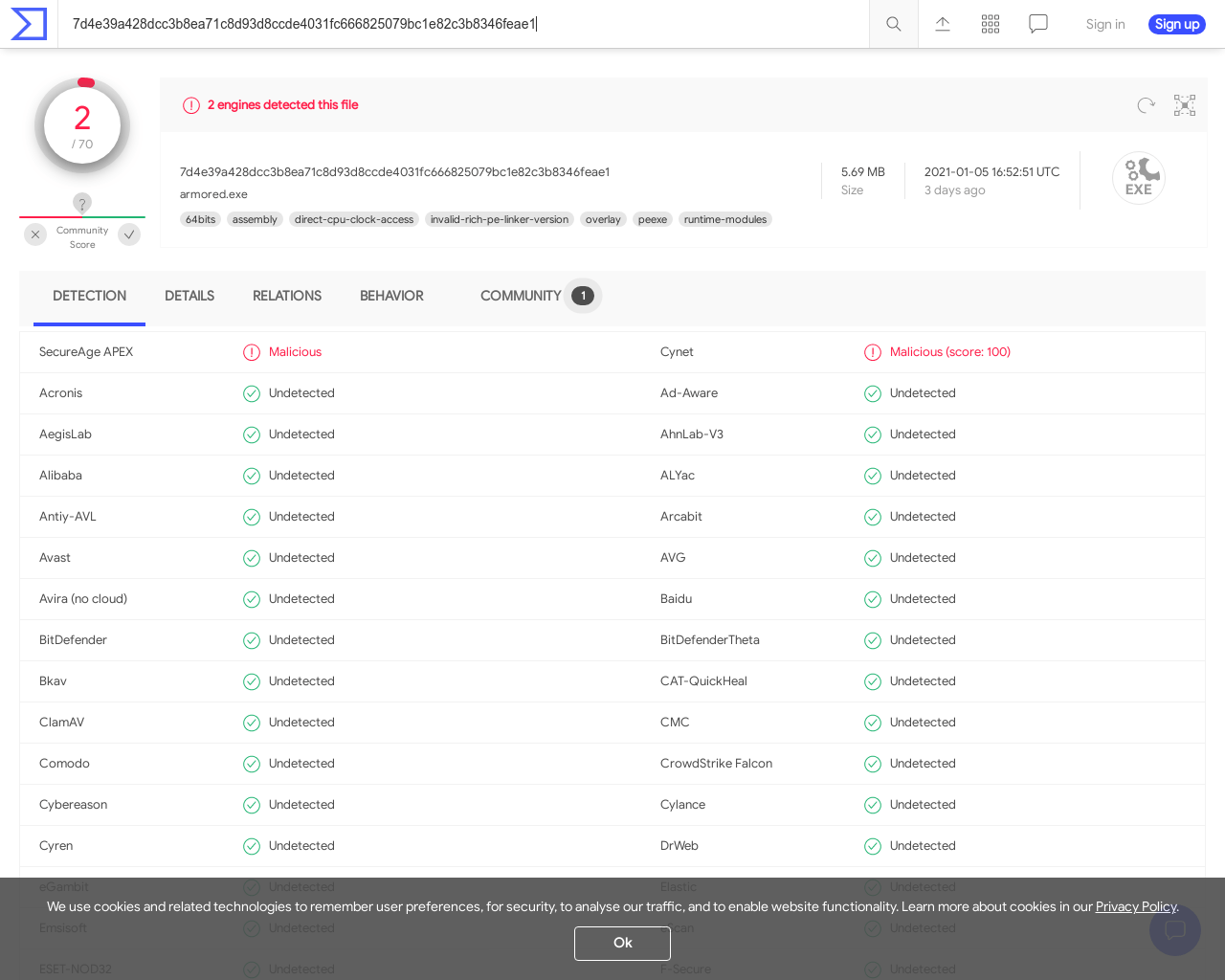}
   \caption{Patched binary with version fix.
   }
   \label{fig:sub5}
 \end{subfigure}~
 \begin{subfigure}[t]{.32\textwidth}
   \centering
   \includegraphics[width=\linewidth]{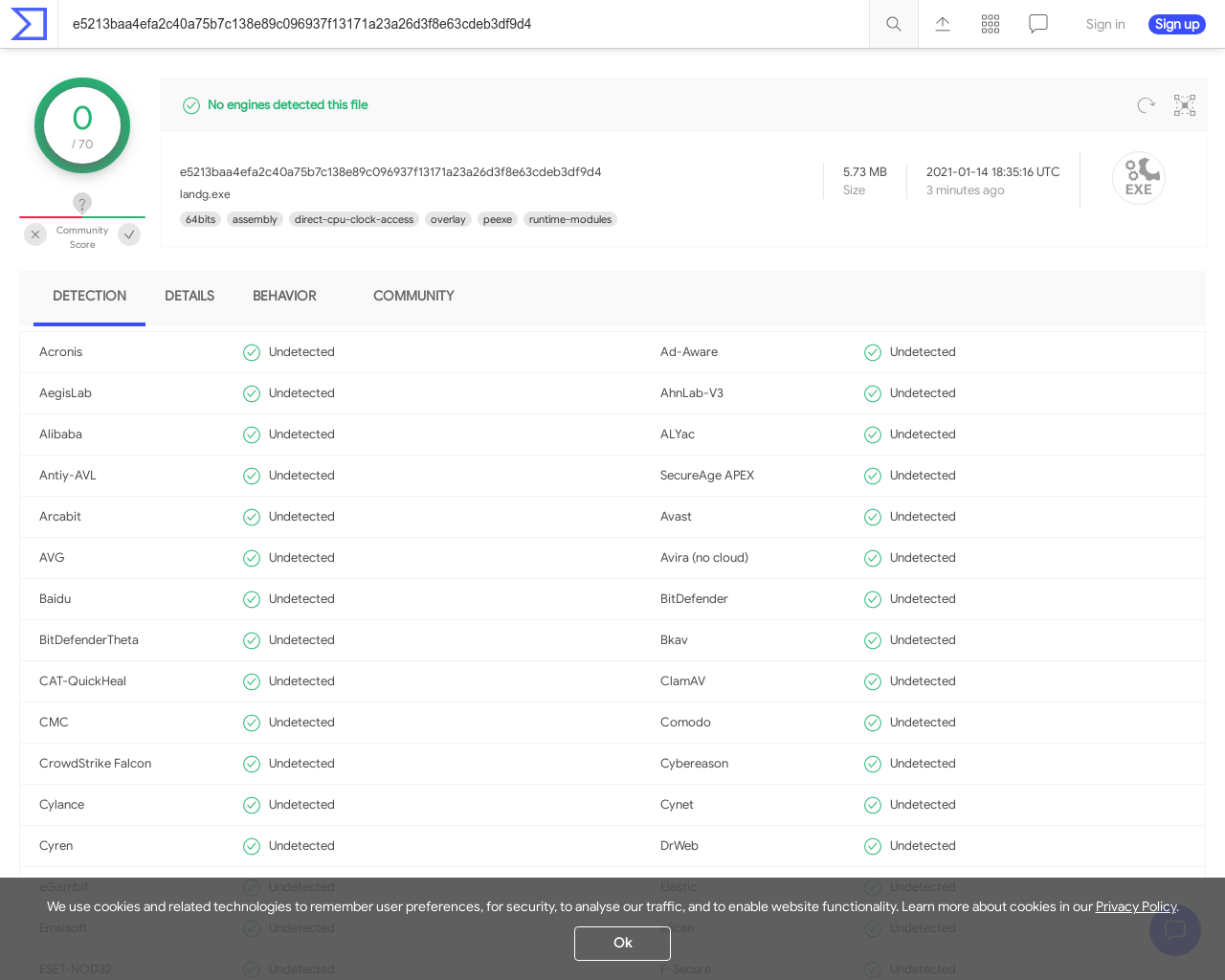}
    \caption{Final binary with all patches.}
    \label{fig:sub6}
 \end{subfigure}

 \begin{subfigure}[t]{.32\textwidth}
  \centering
  \includegraphics[width=\linewidth]{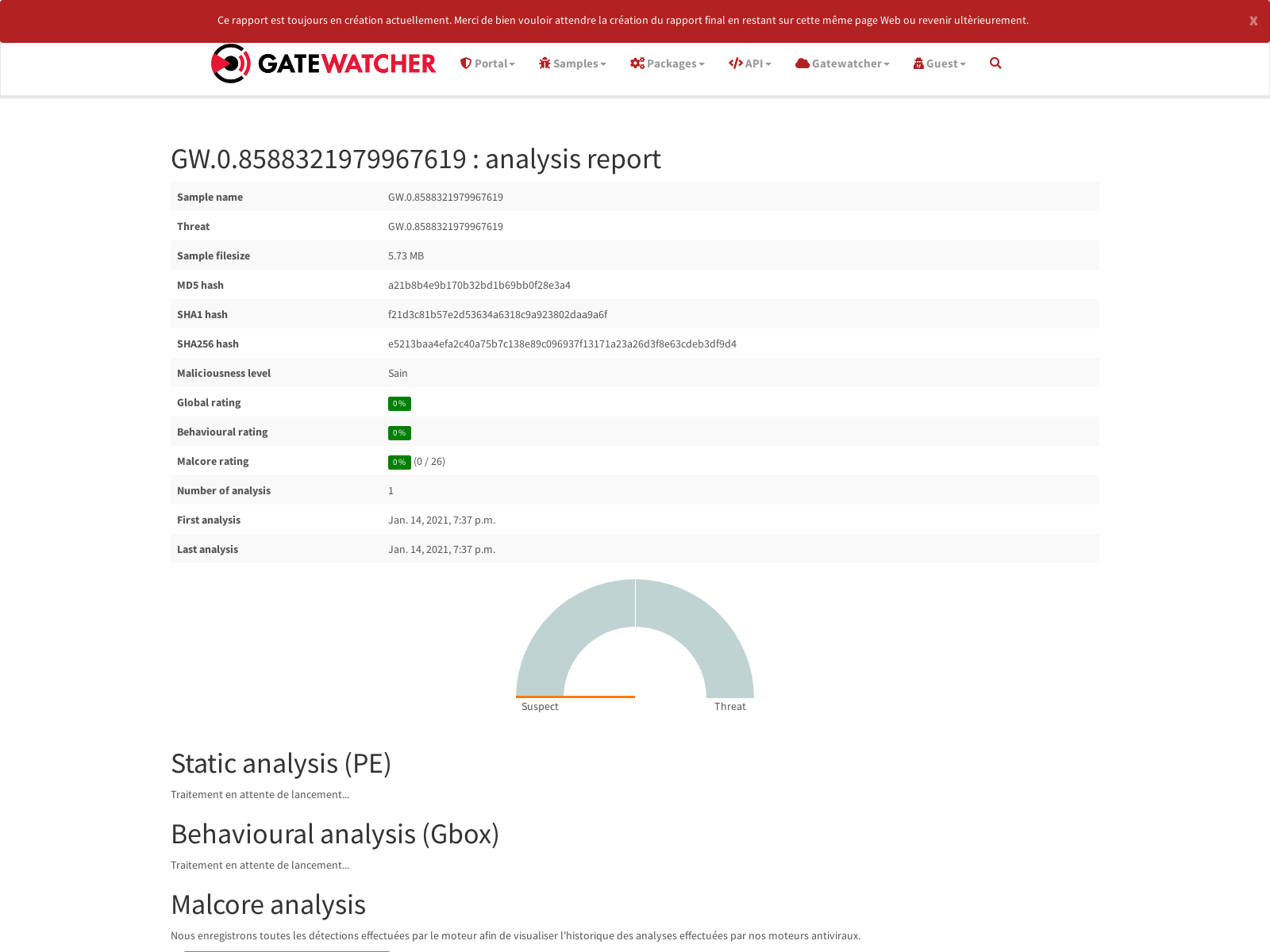}
  \caption{GATEWATCHER scan results.}
  \label{fig:gate}
 \end{subfigure}
  \begin{subfigure}[t]{.32\textwidth}
  \centering
  \includegraphics[width=\linewidth]{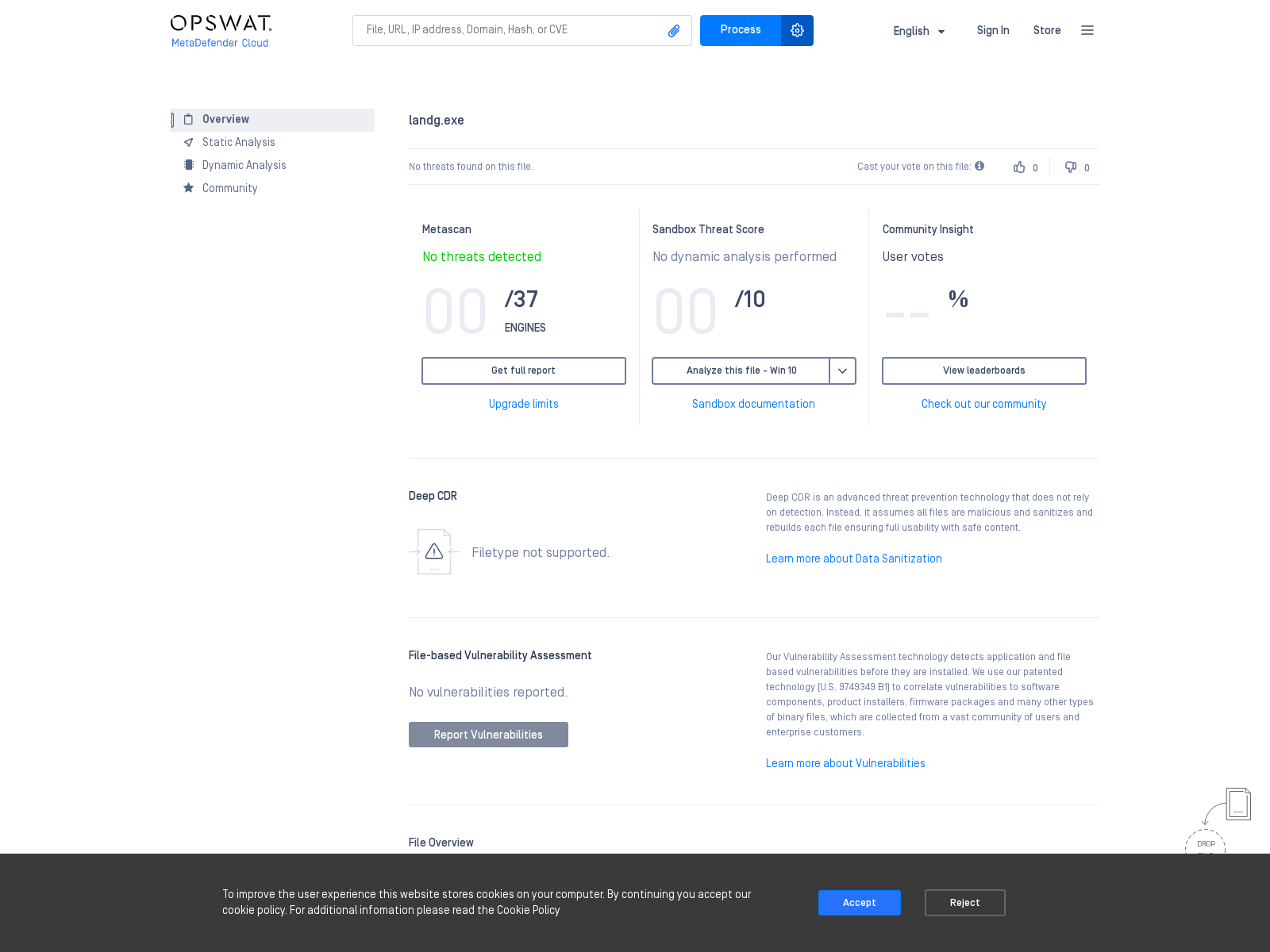}
  \caption{OPSWAT scan results. }
  \label{fig:metadef}
 \end{subfigure}~
 \begin{subfigure}[t]{.32\textwidth}
  \centering
  \includegraphics[width=\linewidth]{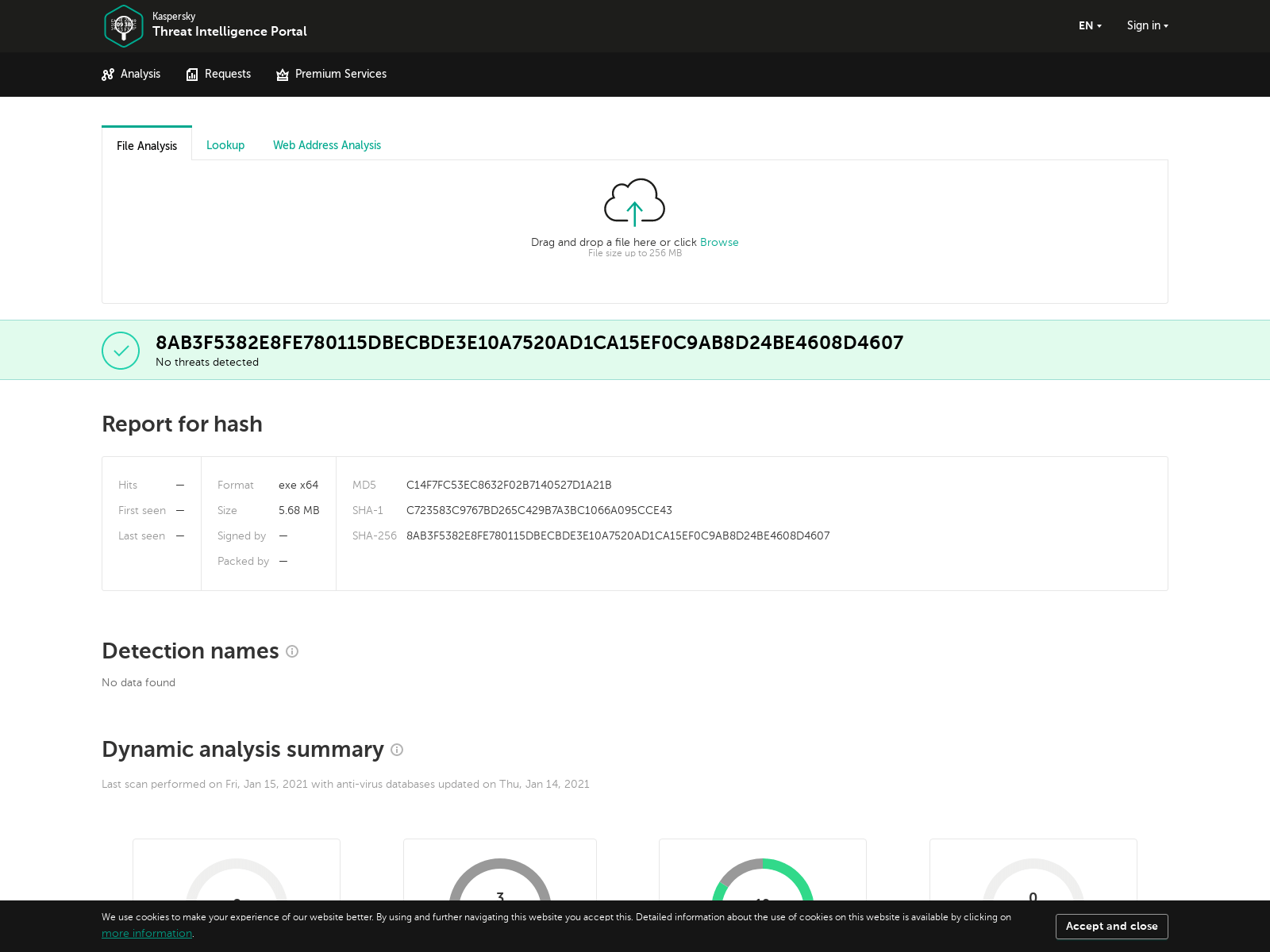}
  \caption{Kaspersky static and dynamic analysis results.}
  \label{fig:kaspersky}
 \end{subfigure}

  \caption{Screenshots of the results of our patched samples from multi-engine scanners.}
  \label{fig:VTResults}
\end{figure*}

Based on the above, it is apparent that by altering the PyInstaller fingerprint on the executable, we may evade the static analyses of many AVs. Thus, to bypass PyInstaller identification by AVs, we initially made some clear ``static'' changes. These changes were i)  substitution of strings and files from ``\textit{pyi\_}'' to a random short string, ii) rename of ``\textit{PyInstaller}'' strings to another random short string, iii) replacement of the default icons, and iv) addition of flags to the linker in WScript, see Table \ref{tbl:flags}. After these changes, we built the new bootloader. We then compiled the malicious script with the modified PyInstaller bootloader, managing to reduce the AVs that reported our executable as malicious to four (Figure \ref{fig:sub4}). Note that the aforementioned actions are bypassing several checks with YARA rules that some AVs might perform, see Figure \ref{fig:yararule}.

 Since our binary did not have any version information, we added one and recompiled it. While a trivial action, after scanning this executable on VT, the AVs are reporting our binary as malicious was further reduced to two (Figure \ref{fig:sub5}). Finally, we opened the last built of our executable with PEtools\footnote{\url{https://github.com/petoolse/petools}}, cleared the rich header and renamed the \texttt{\_RDATA} header to \texttt{.bss} and recalculated the checksum. The removal of the rich header was made to prevent the detection of the binary through the signature of this header \cite{webster2017finding}. This final executable achieved zero detections from VT, see Figure \ref{fig:sub6}. The result was also cross-validated with other custom and multi-engine scanners, e.g. Kaspersky Threat intelligence portal\footnote{\url{https://opentip.kaspersky.com/}}, Gatewatcher\footnote{\url{https://intelligence.gatewatcher.com/}}, MetaDefender\footnote{\url{https://metadefender.opswat.com/}}, see Figure \ref{fig:kaspersky}, \ref{fig:gate} and \ref{fig:metadef}, respectively.

\begin{table}[th]
\centering
\scriptsize
\begin{tabular}{lp{1.5in}}
\toprule
\textbf{Flag} & \textbf{Description}\\ \midrule
/BASE:0x00400000 & Set base to default Windows PE image base  \\
/DYNAMICBASE:NO  & Disable dynamic base                        \\
/VERSION:5.2     & Set image version                           \\
/RELEASE         & Set the checksum of the file                \\ \bottomrule
\end{tabular}
\caption{Linker flags for PyInstaller.}
\label{tbl:flags}
\end{table}

\subsection{Dynamic Analysis with Sandboxes}
To assess the sandboxes and create a proper evasion method, we first need to establish a ground truth baseline for the environment that the sandboxes use. Therefore, the strategy is to initially create a binary that collects intelligence and then aggregate it to make a binary that exploits it to bypass the detection.

To this end, we first created some reconnaissance binaries that were submitted to Intezer, Any.run, Triage, Hybrid Analysis, the public Cuckoo installation of the Estonian CERT\footnote{\url{https://cuckoo.cert.ee/}}, Cape, and Threat Grid sandboxes. However, not all of them allowed Internet connections to the binaries. Therefore, we used a machine with a public IP to collect the input from the reconnaissance binaries when the Internet connection was available. When this was not the case, we manually inspected the logs that were generated from the sandboxes as we wrote the corresponding logs to the disk and registry.

To bypass the execution of our malicious code in a sandbox environment, we analysed the collected data to identify common deficiencies. The most significant misconfiguration in almost all sandboxes was the CPU specifications. More precisely, there were obvious contradictions regarding the threads and cores of the reported CPU. For instance, a sandbox was reporting an AMD EPYC 7371 16-Core Processor, but in the meantime, it was also reporting two cores and two threads. Therefore, we collected all available CPU specifications from Intel and AMD and added them as dictionaries in our the evasive final malware. An aggregated table of the issues that we identified in each sandbox is reported in Table \ref{tbl:sandbox_issues} and will be further discussed in the following paragraphs.

Despite the identified deficiencies, bypassing all of them in a binary is not straight forward. The reason is that continuous calls to read registry values, or WMI is triggering alerts in the sandboxes. Thus, one needs to unify these checks and prioritise them according to the ``noise'' they introduce to the sandbox. Therefore, in our malicious binary, we introduced several conditions before executing the payload.

Firstly, we check whether any known sandbox or VM process is running in the background. Afterwards, we check whether the threads of the system are more than four and if the available RAM is more than 1 GB which is the bare minimum for most of the 64bit modern computers. Then, we check whether the system is powered on more than a threshold, e.g. 2-3 minutes. Next, we examine the foreground applications and the parent of the process of our binary. The reason for this check is the execution process of a sandbox. In most cases, there is a dropper script which opens the file and exits. However, in a real-world execution environment, one would expect that the user would have some other open programs, whether this is the Explorer, Word, or a terminal that would initiate the execution of the binary. Clearly, if this is not the case, then some automated script opened the binary for inspection.

Notably, up to this point, no flag is triggered to the sandbox as the checks do not perform any blacklisted operation and are considered benign by most of them. If all these checks are passed, then we start the registry checks via WMIC for the CPU model name. We validate with our dictionary the existence of the model and the consistency of the reported threads and cores with the manufacturer's actual ones. Usually, this query to the registry is logged by the sandbox, but without any significant alert. Lastly, we query the registry, again via WMIC, to access system information and find known VM strings in the system model or system manufacturer. Clearly, this is also logged by the sandboxes, without though any high score to issue a malicious verdict. Moreover, not all sandboxes managed to reach this point of execution, so in many instances, these logs were not complete in all of the reports.

If any of these checks fail, we perform a graceful exit, perform some arbitrary computations beforehand, and add some noise in the analysis. However, after the successful pass of the aforementioned checks, the malicious binary is executed. Quite alarmingly, in all tested sandboxes, our evasion methods succeeded, achieving low scores in both the dynamic analysis, as well as the static analysis offered by the sandboxes. In fact, all of them considered the samples suspicious for spawning another process of itself which can be considered a false-positive indication, but the malicious payload was not delivered as the binary understood that it was executed in a sandbox.

\section{Discussion}
\label{sec:discussion}
Given the inherent static analysis restrictions, low detection rate from AVs in VT can be considered up to a point expected as our approach is unique and creates an unknown pattern. Nevertheless, the fact that our samples do not simply have few detections, but actually zero is very alarming. It becomes even more worrying because PyInstaller is a widely used tool that is poorly handled. Even the slight changes introduced by us significantly reduced the AVs' detection rate. Notably, these methods can be applied to other languages' packaging, e.g. for Go which is increasingly being used by malware in the past few years\footnote{\url{https://unit42.paloaltonetworks.com/the-gopher-in-the-room-analysis-of-golang-malware-in-the-wild/}}.

It is worth noticing that the above results indicate that AVs do not efficiently handle large executables. For instance, using the UPX feature of PyInstaller to shrink the executable resulted in further detections of the binary. Nevertheless, this can be attributed to the UPX signature. However, the same behaviour was noticed with, e.g. Nuitka\footnote{\url{https://nuitka.net/}} which created far larger executables.

The results of the dynamic sandbox analysis can be considered in many cases, catastrophic. The reason is that our analysis showcases significant issues in the configuration of the sandboxes that allow the malware to fall below their radar. For instance, the vast majority of sandboxes expose inconsistent CPU specifications (processor name vs cores/CPU) while we also noticed the use of non-existing CPU names in one of them. Similar issues were also detected for GPUs.

Differences between CPU timestamp counters may be more challenging to patch; therefore, they were encountered in most sandboxes. Quite interestingly, the listing of well-known VM processes and obvious VM related strings in Bios and system manufacturer (e.g. QEMU, KVM), small uptime, MAC address vendor and low RAM, trivially exposed the virtualisation environment indicating a poor configuration of the sandbox environment. Moreover, we argue that using a limited set of product Windows IDs that we noticed can also be used to fingerprint sandboxes and bypass them. Therefore, the further randomisation of these IDs is necessary as the purchase of more licences does not solve the problem completely.

Finally, we should also stress the complete absence of foreground processes in all sandboxes. In all occasions, the binary started without any other window opened, clearly showing that a dropper initiated the execution. While one may argue that malware may consider this as part of its persistence, e.g. via registry autorun, it would be relatively easy for the malware to verify the claim and correlate it with the uptime. Therefore, sandboxes must open a couple of windows, e.g. Explorer, to denote some user-initiated action for the binary execution and hide the dropper's existence.

\begin{table*}[th]
\centering
\resizebox{.9\textwidth}{!}{
\begin{tabular}{p{1.75in}ccccccc}
& \textbf{Hybrid} &  &  &  &  & \textbf{Thread} &   \\
\textbf{Rules}& \textbf{Analysis} & \textbf{Any.run} & \textbf{Intezer} & \textbf{Tria.ge} & \textbf{Cuckoo} & \textbf{Grid} & \textbf{Cape}  \\
\toprule
Non existing CPU name                             &                 &         &         & \xmark       &        &             &               \\
\rowcolor{lightgray}Bad CPU specifications                            & \xmark               &         & \xmark       & \xmark       & \xmark      & \xmark           & \xmark             \\
Product Key reuse                                 &                 & \xmark       &         & \xmark       & \xmark      &             &               \\
\rowcolor{lightgray}Bad GPU name                                      & \xmark               & \xmark       & \xmark       & \xmark       &        & \xmark           &               \\
Low Available Memory                              & \xmark               &         &         &         &        &             &               \\
\rowcolor{lightgray}Running known VM processes                        &                 & \xmark       &         &         &        &             &               \\
Small uptime                                      &                 &         & \xmark       &         &        &             &               \\
\rowcolor{lightgray}Difference between CPU timestamp counters (rdtsc) & \xmark               &         &         & \xmark       & \xmark      & \xmark           & \xmark           \\
Bad MAC Address                                   & \xmark               &         &         &         &        &             &           \\
\rowcolor{lightgray}Hypervisor bit in CPUId                           &                 &         &         &         & \xmark      &             &           \\
Known VM brand string in Bios                     &                 &         &         &         & \xmark      &             &           \\
\rowcolor{lightgray}Known VM brand string in System Manufacturer      &                 &         &         &         & \xmark      &             &              \\ \bottomrule
\end{tabular}
}
\caption{Identified misconfigurations and issues of sandbox environments.}
\label{tbl:sandbox_issues}
\end{table*}

\section{Conclusions}
In classification, many issues arise from misclassifications and it is essential to understand which features are the ones that resulted to, e.g. a false positive. Based on this problematic, we studied the case of PyInstaller, a widely used packaging tool for Python scripts. The generated executables are erroneously flagged as malicious regardless of their content, as repeatedly reported online by developers. While many malware authors have recently switched to the use of PyInstaller to write their malware, this does not justify why every executable of PyInstaller should be treated as malicious. On the contrary, it implies that AVs do not understand the content of these files and treat them as malicious. Based on this problematic, we have shown that the problem is inherent as AVs cannot efficiently process Python bytecode, which are the \texttt{pyc} files included in PyInstaller. As a result, we may develop malware which escapes static analysis of all AVs by simply changing some characteristics of PyInstaller binaries. Clearly, Python bytecode decompilation is essential to prevent similar attacks in the near future.

Furthermore, based on our analysis, it is evident that apart from clear misconfigurations, resource-wise limitations in the sandboxes impose significant constraints that enable their identification. More precisely, to address the numerous requests for scanning binaries, many of the sandboxes resort to using a limited set of resources (CPU/RAM) which especially for the CPU is not properly handled. As illustrated, many of them report contradictory configurations which can be easily detected and bypassed without issuing any significant alert. The analysis of a binary in a virtualised environment which resembles a traditional, modern PC system is very costly, let alone bear metal analysis. Nevertheless, with the continuous increase of samples that have to be checked, the balance is going to be significantly tipped at the dispense of sandboxes. The latter denotes a definite need to improve our existing sandboxes' capabilities to, e.g. enable them to report more realistic configurations without exposing them. Moreover, we should further explore the analysis using symbolic execution of the binary to offer a cost-efficient alternative. Finally,  despite the recent advances in malware analysis and the numerous academic works and products touting almost absolute detection rates, we illustrate that undetectable malware might even be in plain sight and evade detection in real-world experiments and products.

We argue that one can deploy even stealthier malware by minimising the filesystem footprint. To this end, in future work we plan to rewrite the bootloader to extract all the necessary files in memory or use PyOxidizer\footnote{\url{https://github.com/indygreg/PyOxidizer}}, randomising file names in each compilation, further reducing the pattern that one could use to trace it. Fileless approaches \cite{kumar2020emerging} in which all the content is loaded in memory through the use of, e.g. Living Off The Land Binaries And Scripts (LOLBins and LOLScripts)\footnote{\url{https://github.com/LOLBAS-Project/LOLBAS}} can further decrease the detectability of the binary. In parallel, we plan to investigate other packaging and distribution tools for other languages beyond Python to assess their obfuscation abilities.

\section*{Acknowledgements}
This work was supported by the European Commission under the Horizon 2020 Programme (H2020), as part of the projects CyberSec4Europe (\url{https://www.cybersec4europe.eu}) (Grant Agreement no. 830929), \textit{LOCARD} (\url{https://locard.eu}) (Grant Agreement no. 832735).

The content of this article does not reflect the official opinion of the European Union. Responsibility for the information and views expressed therein lies entirely with the authors.

\bibliographystyle{plain}

\bibliography{references}

\end{document}